\documentclass[]{aastex62}
 
\usepackage{natbib}
\shorttitle{Semi-analytic simulations of First Stars}
\shortauthors{Visbal et al.}

\begin{document}

\title{Self-consistent Semi-analytic Modeling of Feedback During Primordial Star Formation and Reionization}

\email{Elijah.Visbal@utoledo.edu}

\author{Eli Visbal}
\affil{University of Toledo, Department of Physics and Astronomy, 2801 W. Bancroft Street, Toledo, OH, 43606, USA}
\affil{Ritter Astrophysical Research Center, 2801 W. Bancroft Street, Toledo, OH 43606, USA}

\author{Greg L. Bryan}
\affil{Center for Computational Astrophysics, Flatiron Institute, 162 5th Ave, New York, NY, 10003, USA}
\affil{Columbia University, Department of Astronomy, 550 West 120th Street, New York, NY, 10027, USA}

\author{Zolt\'{a}n Haiman}
\affil{Columbia University, Department of Astronomy, 550 West 120th Street, New York, NY, 10027, USA}

\begin{abstract} 
We present a new semi-analytic model of the formation of the first stars.
Our method takes dark matter halo merger trees (including 3-dimensional spatial information) from cosmological N-body simulations as input
 and applies analytic prescriptions to compute both the Population III (Pop III) and metal-enriched star formation histories.
We have developed a novel method to accurately compute 
the major feedback processes affecting Pop III star formation: H$_2$ photodissociation from 
Lyman-Werner (LW) radiation, suppression of star formation due to inhomogeneous reionization, 
and metal enrichment via supernovae winds. Our method utilizes a grid-based 
approach relying on fast Fourier transforms (FFTs) to rapidly track the LW intensity, ionization fraction,
and metallicity in 3-dimensions throughout the simulation box. 
We present simulations for a wide range of astrophysical model parameters from $z\approx 30-6$.
Initially long-range LW feedback and local metal enrichment and reionization feedback 
dominate. However, for $z \lesssim 15$ we find that 
the star formation rate density (SFRD) of Pop III stars is impacted
by the combination of external
metal enrichment (metals from one halo polluting other pristine halos) and inhomogeneous reionization.
We find that the interplay of these processes is particularly important for the Pop III SFRD at $z \lesssim 10$.
Reionization feedback delays star formation long enough for metal bubbles to reach halos that would 
otherwise form Pop III stars. Including these effects can lead to more than 
an order of magnitude decrease in the Pop III SFRD at $z=6$ compared to LW feedback alone.
\end{abstract}

\keywords{stars: Population III – galaxies: high-redshift – cosmology: theory}

\section{Introduction}
Simulations based on the standard model of cosmology predict that the first Pop III stars 
formed within $\sim100$ Myr after the Big Bang in small $\sim10^{5}~M_\odot$ dark matter ``minihalos'' \citep[for a recent review see][]{2015ComAC...2....3G}.
Currently, there are no definitive observations of Pop III stars, but a variety of upcoming facilities are positioned 
to improve constraints and may lead to unambiguous observational evidence. 
Radio telescopes targeting the cosmological 21cm signal \citep{2006PhR...433..181F, 2012RPPh...75h6901P}, both the global sky-averaged spectrum
\citep[e.g., \emph{EDGES}  and  \emph{LEDA},][]{2018Natur.555...67B, 2018MNRAS.478.4193P}
and spatial fluctuations \citep[e.g.,~\emph{HERA} and \emph{SKA},][]{2017PASP..129d5001D, 2015aska.confE...1K},
are able to study the first stars through their impact on the intergalactic medium (IGM) \citep[e.g.,][]{2012Natur.487...70V, 2013MNRAS.432.2909F, 2017MNRAS.472.1915C, 2018Natur.555...67B}. Observations of extremely metal poor
stars in the local Universe, which will be rapidly identified with the next generation of 30-meter class telescopes, can constrain 
the properties of the first stars \citep[e.g.,][]{2015MNRAS.447.3892H, 2018MNRAS.478.1795H,2018MNRAS.473.5308M}. 
Pop III stars in the 
mass range of $\sim 140-250~M_\odot$ may end their lives as extremely bright pair instability supernovae, which can potentially 
be detected with the \emph{James Webb Space Telescope} (\emph{JWST}) \citep{2013ApJ...762L...6W, 2018MNRAS.479.2202H}. 
Additionally, helium recombination line-intensity mapping may be able to constrain the global star formation history of Pop III stars \citep{2015MNRAS.450.2506V} and
it is possible to detect the impact of the first stars on the ionization fraction of the high-redshift IGM
with cosmic microwave background (CMB) measurements \citep[e.g.,][]{2003ApJ...583...24K, 2003ApJ...595....1H,  2012ApJ...756L..16A, 2015MNRAS.453.4456V, 2017MNRAS.467.4050M}.

In order to make accurate predictions for these observational probes, it is necessary to model the 
abundance of Pop III stars as a function of time. This requires accounting for a number of important feedback processes 
including metal enrichment from supernovae winds, suppression of star formation from hydrogen ionizing radiation, and LW photodissociating radiation, 
which suppresses or delays Pop III star formation in small dark matter halos \citep{1997ApJ...476..458H, 2001ApJ...548..509M, 2008ApJ...673...14O}.
A key challenge is that these processes operate over a vast range of distance scales.
For example, metal enrichment due to Pop III supernovae explosions
acts within individual dark matter halos or nearby regions of the IGM ($\lesssim 100$ comoving kpc). On the other hand,
 LW feedback is due to the total LW flux from stars within ${\sim} 100 $ Mpc of any potential star-forming halo \citep{2009ApJ...695.1430A}. We note
 that X-ray feedback due to gas accretion onto black hole remnants of primordial stars may operate on even larger scales \citep{2004ApJ...604..484M, 2004MNRAS.352..547R, 2005MNRAS.357..207R, 2012MNRAS.425.2974T}, however we do not include black hole modeling and X-ray feedback in this paper. We defer its treatment for future work.
 
Several different theoretical approaches have been utilized to make predictions for the abundance of Pop III stars. 
The most detailed calculations are hydrodynamical cosmological simulations including star formation,
radiative feedback, and metal enrichment via supernovae winds \citep[e.g.,][]{2012MNRAS.427..311W, 2012ApJ...745...50W, 2014MNRAS.442.2560W, 2015ApJ...807L..12O, 2016ApJ...823..140X}. 
Simulations permit the most accurate treatment of the relevant physical processes, but are computationally expensive \citep[e.g.,~tens of millions of CPU-hours for the Renaissance simulations][]{2015ApJ...807L..12O, 2016ApJ...823..140X}. 
Despite their sophistication, numerical simulations still require subgrid 
physics which is put in by hand. This includes the Pop III IMF, the metal-enriched IMF, 
and the critical metallicity for the Pop III/Pop II transition.
The numerical demands restrict
hydrodynamical cosmological simulations to relatively small ($\sim 10$ Mpc) simulation boxes (or zoom-ins) and limit the number of runs which can be performed to explore the parameter space of sub-grid physics.

The other extreme, in terms of theoretical approaches, is to use analytic calculations which can rapidly model large effective volumes and 
the uncertain parameter space of the high-redshift Universe. The simplest treatments
involve estimating a minimum mass of dark matter halo that hosts Pop III stars and integrating over the halo mass function \citep[e.g.,][]{2005MNRAS.361..577C, 2005ApJ...623....1M, 2007ApJ...659..890W, 2009ApJ...694..879T,2006ApJ...650....7H, 2015MNRAS.453.4456V}.
These calculations are numerically inexpensive, but typically lack 3-dimensional spatial information and halo merger histories
 which are important for feedback and observational predictions.  Additionally, halo merger histories can be modeled rapidly with Monte Carlo methods \citep[as in][]{2016MNRAS.462.3591M}, but this approach still lacks 
3-dimensional spatial information.
We note that in some instances, 3-dimensional clustering has been modeled analytically \citep{2002ApJ...571..585S, 2003ApJ...589...35S, 2006ApJ...649..570K}. In \cite{2003ApJ...589...35S}, an analytic treatment of clustering was utilized to determine the abundance of Pop III stars as a function of cosmic time, including the impact of metal pollution due to supernovae winds. However, in contrast to the models presented here, they do not take into account the impact of cosmic reionization and the baryon-dark matter streaming velocities \citep{2010PhRvD..82h3520T}. \cite{2006ApJ...649..570K} have incorporated photo-ionization feedback into a toy-model that approximates source clustering, but only in the radial direction away from sources, whereas here we capture the full three-dimensional clustered distribution of the ionizing sources.

In this paper, we focus on a hybrid method between the numerical and analytic approaches just described. 
This ``semi-analytic'' method utilizes dark matter-only cosmological N-body simulations to determine the
spatial and merger properties of dark matter halos. Analytic prescriptions are then applied to the 
dark matter merger trees to track both Pop III and metal-enriched star formation. 
Semi-analytic models allow much faster computations compared to hydrodynamical simulations (each run of our model presented below takes roughly 24 CPU-hours)
and can incorporate more detailed physics than purely analytic methods. The computational efficiency of our method makes rapid parameter space exploration possible. This method can also be used to predict the scatter in observables due to random changes in initial conditions (cosmic variance). 

A number of previous works have utilized semi-analytic simulations to study the 
first stars and galaxies \citep[e.g.,][]{2009ApJ...700.1672T, 2012MNRAS.425.2854A, 2013ApJ...773..108C, 2016MNRAS.457.3356V, 2018MNRAS.479.4544M, 2018MNRAS.475.5246V, 2019arXiv191010171M}. 
Some of these works have focused on predictions for Pop III or second generation stars which could be found in the Milky Way at $z=0$ \citep[e.g.,][]{2016arXiv161100759G, 2016ApJ...826....9I, 2017MNRAS.465..926D, 2017MNRAS.469.1101G, 2018MNRAS.473.5308M}.
There have also been efforts to model galaxies during
the epoch of reionization without following Pop III stars or minihalos \citep{2016MNRAS.459.3025P, 2016MNRAS.462..250M, 2019MNRAS.483.2983Y, 2019MNRAS.490.2855Y}. We also note that semi-analytic methods have been very useful in the context of more evolved galaxies at low redshift \citep[e.g.,][]{2006RPPh...69.3101B, 2014ApJ...795..123L}.

In addition to N-body simulations or Monte Carlo merger trees, semi-analytic models can be created using dark matter halos simulated with a code such as {\sc pinocchio} \citep{2002ApJ...564....8M}, which utilizes Lagrangian Perturbation Theory (LPT) to rapidly model the formation and properties of halos in cosmological volumes. This has been applied to the problem of Pop III star formation in \cite{2019MNRAS.483.3592B} (though they use a much simpler model of feedback processes than the present work). The LPT and N-body methods are complementary in the sense that LPT can generate larger cosmological volumes at low computational expense, but N-body simulations can more accurately capture the small-scale properties of halos (e.g., clustering on non-linear scales). In future work, a detailed side-by-side comparison would be helpful to determine the exact limitations of the LPT approach in the context of the first stars.

The main new feature of our model presented below is a grid-based approach to 
calculate the major feedback processes for Pop III star formation: LW feedback, suppression of star formation
due to photoheating from reionization, and metal enrichment due to supernovae winds.
Our grid-based method utilizes FFTs
to rapidly compute the 3-dimensional LW intensity, ionization fraction, and metallicity throughout our
simulation box. 
We note that these models are also the first of their type
to include a treatment of the baryon-dark matter velocities \citep{2010PhRvD..82h3520T}, which suppress star formation in small dark matter
halos in high-velocity regions \citep{2012MNRAS.424.1335F}.

We utilize this method to compute the Pop III and metal-enriched star formation histories across cosmic time for
a range of the uncertain model parameters describing star formation and feedback. Detailed results are 
presented below. We find that spatial clustering of feedback effects begins to be important at $z \lesssim 15$,
when a significant fraction of the IGM is ionized. We also find that at high redshifts, star formation 
is strongly impacted by the baryon-dark matter streaming velocity. 
Another important conclusion is that the timing of the cosmic transition between Pop III and metal-enriched star formation 
is likely set by the delay between initial metal-free star formation and second generation star formation due
to supernovae explosions of Pop III stars.

This paper is structured as follows. In Section 2, we describe the cosmological simulations utilized by our model. We 
explain the details of our model in Section 3 and present results for the Pop III and metal-enriched cosmic star formation 
histories for a range of model parameters in Section 4. Finally, we discuss our results and main conclusions in Section 5.
Throughout, we assume a $\Lambda {\rm CDM}$ cosmology with parameters consistent with \cite{2014A&A...571A..16P}: $\Omega_{\rm m} = 0.32$, $\Omega_{\Lambda} = 0.68$, $\Omega_{\rm b} = 0.049$, $h=0.67$, $\sigma_8=0.83$, and $n_{\rm s} = 0.96$.

\section{N-Body Simulations}
We utilize the publicly available code {\sc gadget2} \citep{2001NewA....6...79S} to produce the cosmological N-body 
simulations required for our semi-analytic model. To improve the statistics of our results, we ran 10 simulations, each with a different random seed for the initial conditions. All of these simulations have a 3 Mpc box length and $512^3$ particles.  This corresponds to a particle mass of $8\times10^3~M_\odot$.  
Initial conditions were generated with {\sc2LPTIC} \citep{2006MNRAS.373..369C} and created
 at $z=200$. For each run, $\sim100$ snapshots were saved from $z=40$ to $z=6$, spaced in cosmic time by 
$\Delta t = t_{\rm H}/40$, where $t_{\rm H}$ is the Hubble time at the preceding snapshot. This time spacing corresponds to
$\approx 0.25$ of the dynamical time of a halo at its virial radius, which we find is sufficient for convergence in our semi-analytic model.
Halo catalogs and merger trees were computed using the publicly available codes {\sc rockstar} \citep{2013ApJ...762..109B} and {\sc consistent trees} \citep{2013ApJ...763...18B}. We emphasize that the selected mass resolution is sufficient for our semi-analytic models. As discussed below (see Figure \ref{popIII_masses}), the first Pop III stars in our model form in $\sim 2\times 10^6~M_\odot$ dark matter halos. These are resolved with $\sim 250$ dark matter particles in our N-body simulations. {\sc rockstar} has been shown to accurately determine halo properties for halos with $\gtrsim 20$ particles \citep{2013ApJ...762..109B, 2011MNRAS.415.2293K}. Previous work has also shown that substructure in dark matter halos can be tracked for subhalos with $\gtrsim 100$ particles \citep{2012MNRAS.423.1200O}. We do not attempt to track substructure in our model, leading to a less stringent requirement. Thus, we regard our 250 particle resolution as a relatively conservative choice.
We also note that the simulated halo mass functions agree very well with the analytic mass function of  \cite{1999MNRAS.308..119S} \citep[see Figure 1 in][]{2018MNRAS.475.5246V}.

\begin{table*}
\centering
\caption{\label{table} Key physical parameters entering the semi-analytic model, their fiducial values, and the ranges we have varied them. Except where otherwise noted in parentheses, these are defined in Section 3.1.}
\begin{tabular}{c l| c || c }
Parameter & Description & Fiducial Value  & Range \\ 
\hline
$f_{\rm III}$ & Pop III star formation efficiency  &  0.001 & $ 0.0001- 0.005$ \\[1ex]
$f_{\rm II}$ & Metal-enriched star formation efficiency &  0.05 &  -- \\[1ex]
$\eta_{\rm II}$ & LW/Ionizing photons per baryon of metal-enriched stars &  4000 & -- \\[1ex]
$\eta_{\rm III}$ & LW/Ionizing photons per baryon of Pop III stars &  65000 & -- \\[1ex]
$t_{\rm delay}$ & Delay in subsequent star formation due to Pop III SNe &  $10^7$ yr & $10^7 {\rm yr} -  5\times 10^7 {\rm yr}$ \\[1ex]
$Z_{\rm crit} $ & Critical metallicity for externally metal-enriched halos  &  $3 \times 10^{-4}~Z_\odot$ &  $10^{-6}~Z_\odot - 10^{-2}~Z_\odot$\\[1ex]
$M_{\rm min, met} $ & Critical mass for externally metal-enriched halos  &  $2\times 10^5~M_\odot$ &   $2\times 10^5~M_\odot -10^6~M_\odot$ \\[1ex]
$M_{\rm ion} $ & Ionization feedback mass &  $1.5\times 10^8  \left ( \frac{1+z}{11} \right )^{-3/2}~M_\odot$  &  $ {\rm Fiducial} - 3.3 \times {\rm Fiducial}$ \\[1ex]
$f_{ \rm esc, II}$ & Ionizing escape fraction in metal-enriched halos  (Section 3.3) &  0.1 & $0.0-0.1$  \\[1ex]
$f_{ \rm esc, III}$ & Ionizing escape fraction in Pop III halos (Section 3.3) &  0.5 & $0.0-0.5$  \\[1ex]
$v_{\rm bc}$ & Streaming velocity at recombination &  $30~{\rm km~s^{-1}} ~ (1\sigma)$ &  $0-90~{\rm km~s^{-1}}$ ($0-3\sigma$)  \\[1ex]
$f_{\rm bub}$ & Metal bubble size scaling factor (Section 3.4) &  1 & $0-5$ \\[1ex]
$M_{\rm min}$ &  Minimum halo mass for Pop III star formation   &   see Eqns. 1-3   &   -- \\[1ex]
\end{tabular}
\end{table*}

\section{Semi-analytic Model}
Broadly speaking, our semi-analytic model takes as input the merger history and 
3-dimensional spatial information of dark matter halos from an N-body simulation.
It then utilizes analytical prescriptions to determine where, when, and how much Pop III and metal-enriched 
star formation occur. These prescriptions include major feedback processes
relevant for Pop III star formation: LW feedback, reionization of the IGM, and metal enrichment from 
supernovae winds. This physics is characterized by a number of model parameters which 
are listed with their fiducial values in Table \ref{table}.

The most important novel feature of the model
is a grid-based method for computing feedback processes which 
depend on the 3-dimensional positions and clustering of dark matter halos. As described in the following subsections,
this method utilizes FFTs to rapidly compute the local values of the LW intensity as well as the 
metallicity and neutral fraction of the IGM throughout the simulation box.  
The new method allows us, for the first time, to rapidly compute HII regions due to ionizing photons from Pop III stars and metal-enriched galaxies
including the impact of multiple sources inside ionization bubbles \citep[this was inspired by work on larger distance scales, such as] []{2004ApJ...613....1F}. This is not properly accounted for in simpler `shell' models
that have been employed in previous semi-analytic simulations tracking Pop III star formation \citep[e.g.,][]{2018MNRAS.473.5308M, 2018MNRAS.475.5246V}. A shell approach which computes an ionized bubble radius for each source separately and
does not take into account how nearby ionizing sources both ionize some of the same portion of the IGM, underestimates the total ionized volume around these clustered sources.

\subsection{Star formation}
Dark matter halos throughout the simulation box are assigned both Pop III and metal-enriched star formation.
Pop III star formation occurs when a pristine halo (i.e.~no external/internal metal enrichment as described below)
first reaches the minimum mass where gas can cool and form stars.
For metal-free minihalos, this cooling is due to molecular hydrogen transitions.
Hydrogen molecules are photo-dissociated by LW radiation, which can suppress Pop III star formation in smaller minihalos
\citep{1997ApJ...476..458H,2001ApJ...548..509M,2007ApJ...671.1559W,2008ApJ...673...14O,2011MNRAS.418..838W,2014MNRAS.445..107V}.
The baryon-dark matter streaming velocity  \citep{2010PhRvD..82h3520T} can also suppress star formation in small minihalos \citep{2012MNRAS.424.1335F}.
Considering both of these effects, we assume that for halos
which are not in an ionized region of the IGM, the minimum mass for Pop III star formation is
\begin{equation}
M_{\rm min} = \min \left ( M_{\rm H2},  M_{\rm a} \right ),
\end{equation}
where 
\begin{equation}
\label{H2_eqn}
M_{\rm H2} =  M_{\rm cool}(v_{\rm bc}, z) \times  \left (1 + 6.96[4\pi J_{\rm LW, 21}]^{0.47} \right)
\end{equation}
is the minimum mass for molecular cooling in minihalos \citep[][]{2013MNRAS.432.2909F} and $M_{\rm a}= 5.4 \times 10^7 ([1+z]/11)^{-3/2}$ is the atomic cooling threshold found in hydrodynamical cosmological simulations \citep[with the precise value taken from the simulations of][]{2014MNRAS.439.3798F}. The local value of the LW background, 
$J_{\rm LW, 21}$, is given in units of $10^{-21}~{\rm erg ~ s^{-1} ~ cm^{-2}  Hz^{-1} ~  sr^{-1}}$.
Here $M_{\rm cool}$ is the cooling mass for a minihalo without LW radiation as a function of redshift and the local baryon-dark matter streaming velocity,
$v_{\rm bc}$. Note that the value of the streaming velocity evolves as $v_{\rm bc} \propto (1+z)$ and has a typical value of 30 ${\rm km~s^{-1}}$  (the root-mean-square value which we denote as $\sigma$) at recombination. We compute $M_{\rm cool}$ from the following fitting function calibrated to hydrodynamical simulations by \cite{2012MNRAS.424.1335F}, 
\begin{equation}
V_{\rm cool}^2 = \left ( a^2 + [b v_{\rm bc}(z)]^2  \right )^{1/2}.
\end{equation}
In this formula, $V_{\rm cool}$ is the circular velocity corresponding to the viral mass $M_{\rm cool}$ \cite[as defined in][]{2001PhR...349..125B} and the fit parameters are $a=3.714~{\rm km/s}$ and $b=4.015$ \citep[which are determined from the simulations of][]{2011ApJ...736..147G, 2011ApJ...730L...1S}.
On scales of $\sim 3 ~{\rm Mpc}$ the streaming velocity is roughly constant,
so for any particular run, we assume one value of the streaming velocity. We note that because 
our N-body simulations are dark matter only, we are not able to incorporate the reduction in the halo mass function 
due to high $v_{\rm bc}$ \citep{2010PhRvD..82h3520T}. However, this is a modest effect \citep[e.g., $\sim$20 percent reduction of $5\times 10^4~M_\odot - 5\times 10^5~M_\odot$ halos at $z=25$ for $v_{\rm bc}= \sigma$,][]{2012ApJ...747..128N} and should be secondary compared to the 
delay in molecular cooling as described in \cite{2013MNRAS.432.2909F}.
Future work could use hydrodynamical cosmological simulations to compute merger trees with the streaming velocity included.
We note that the fit in Eq. \ref{H2_eqn} is based on a relatively small number of hydrodynamical simulations. Future work exploring a larger range 
of redshifts and combining streaming velocities and LW feedback will be important to accurately tune models like those presented in this paper.

When a pristine halo reaches $M_{\rm min}$, we assume that Pop III stars are formed with 
an efficiency $f_{\rm III}$ (i.e.~a total mass $M_{\rm *, III} = f_{\rm III}M_{\rm h} \Omega_{\rm b}/ \Omega_{\rm m}$ of Pop III 
stars are formed). We also assume that there is then a delay, $t_{\rm delay}$, due to the lifetime of 
the stars and recovery from SN explosions before
metal-enriched star formation can occur. The recovery time is needed for gas to resettle in a halo's gravitational potential well after being disrupted by SNae.  We assume a fiducial value of $t_{\rm delay}=10~{\rm Myr}$, which 
 represents recovery from a core collapse SN \citep{2014MNRAS.444.3288J} in a $\sim 5\times 10^5~M_\odot$ minihalo. The delay in efficient metal mixing could take longer in more massive dark matter halos. The impact of increasing $t_{\rm delay}$ is discussed in detail below (which is significant for the timing of the cosmic transition from Pop III to metal-enriched star formation).
Once a time greater than $t_{\rm delay}$ has passed in a halo (starting from the earliest Pop III episode if there are multiple Pop III progenitors),
metal-enriched star formation begins. We assume that across each time step in our simulations (which are the same as the N-body snapshots, set to $\approx 0.25$ of the dynamical time of a dark matter halo at the virial radius as described in Section 2), the metal-enriched star formation rate is
\begin{equation}
\label{SF_II}
{\rm SFR_{\rm II}} = \frac{f_{\rm II} M_{\rm accrete}\Omega_{\rm b}}{\Omega_{\rm m} \Delta t},
\end{equation}
where $M_{\rm accrete}$ is the mass accreted from halos without stars
or in smooth accretion (including halos below the resolution of the simulation) and $\Delta t$ is the time between N-body simulation snapshots. 
Thus, a minihalo which hosts Pop III star formation must accrete additional material or merge with other pristine minihalos before it can 
form a significant amount of metal-enriched stars in our model. We find that this typically happens over our assumed values of $t_{\rm delay}$.
In our fiducial model, we have adopted a metal enriched star formation efficiency of $f_{\rm II}=0.05$, which we take from \cite{2015MNRAS.453.4456V}. This value was determined using abundance matching and the observed UV luminosity function of galaxies at $z\approx 6$ from \cite{2015ApJ...803...34B}. 
We note that the Pop III star formation efficiency is not tightly constrained by simulations. We have adopted a fiducial value of $f_{\rm III} = 0.001$, 
which corresponds to $\sim 100~M_\odot$ of stars in a $\sim 10^6~M_\odot$ minihalo and is compatible with the CMB optical depth 
due to electron scattering with our fiducial set of model parameters \citep{2015MNRAS.453.4456V}. 

In addition to internal enrichment (i.e.~star formation and enrichment in the same halo as just described), we also include 
external enrichment (supernovae winds from one halo enriching another separate halo). As discussed below,
we use our grid-based approach to track the metallicity of the IGM and if a halo with mass above $M_{\rm min, met}$ is in a region with metallicity above 
$Z_{\rm crit}$, we assume that metal-enriched star formation proceeds according to Eq. \ref{SF_II}. 
We assume a fiducial value of $M_{\rm min, met} = 2 \times 10^5 ~ M_\odot$, but note that this is a highly uncertain quantity. 
For $Z_{\rm crit}$, we assume a fiducial value of $3\times 10^{-4}~Z_\odot$. There have been a number of previous studies 
attempting to determine this quantity  \citep[e.g.,][]{2003Natur.425..812B, 2005ApJ...626..627O, 2007ApJ...661L...5S, 2009ApJ...691..441S}. 
These works tend to find a critical value near our fiducial choice, but it could be significantly lower
 ($Z_{\rm crit} \approx  10^{-6}~Z_\odot$) if dust cooling is important, rather than just C and O \citep{2005ApJ...626..627O}. 

We also include the impact of reionization of the IGM. The neutral fraction of the simulation box is tracked
with the grid-based method as described below (assumed to be 0 or 1 in each resolution element of our grid). 
Gas heating due to reionization has been shown to impede star formation in smaller dark matter halos 
\citep[e.g.,][]{1994ApJ...427...25S,1996ApJ...465..608T,1998MNRAS.296...44G,2000ApJ...542..535G,2004ApJ...601..666D,2006MNRAS.371..401H,2008MNRAS.390..920O,2013MNRAS.432L..51S,2014MNRAS.444..503N}.
In our model, if a halo without prior star formation is in an ionized region, 
the minimum mass is raised (in the fiducial model) to $M_{\rm min, ion} = 1.5\times 10^8  \left ( \frac{1+z}{11} \right )^{-3/2}~M_\odot$ \citep[consistent with][]{2004ApJ...601..666D}.
For a halo in an ionized region which already has star formation, the metal-enriched star formation is reduced by a factor of 
$\exp (-t_{\rm rei}/[0.1t_{\rm H}])$, where $t_{\rm rei}$ is the time since reionization for the first progenitor halo which was reionized.
This factor's characteristic timescale is equal to the dynamical time at the virial radius of a recently formed dark matter halo, $t_{\rm dyn} \approx 0.1t_{\rm H}$.
We include this timescale to account for the fact that star formation may not be instantaneously quenched when the IGM surrounding a galaxy 
is ionized. Dense gas in the halo which is already flowing towards the center may not be strongly impacted even if lower density gas in the surrounding
IGM (outside of the halo) is ionized.

Our LW and ionizing feedback prescriptions, described in detail below, depend on the amount of LW and ionizing radiation produced by stars.
We assume that the number of hydrogen ionizing photons produced per baryon incorporated into stars is
$\eta_{\rm II}= 4000$ and $\eta_{\rm III}=65000$, for metal-enriched and Pop III stars, respectively. The metal enriched value
corresponds to a Salpeter IMF from 0.1 to 100 $M_\odot$ and metallicity $Z=0.0004$ \citep[see table 1 in][]{2007MNRAS.377..285S}. 
The Pop III value is expected for a $\sim 40~M_\odot$ star over its lifetime \citep{2002A&A...382...28S}. For simplicity, we assume one LW 
photon is created for each ionizing photon. This assumption may underestimate the LW flux by a factor of a few for
metal-enriched galaxies. Given the relatively large uncertainties in a number of astrophysical parameters (e.g., the star formation efficiencies),
we leave a more detailed treatment for future work. We also note that we have assumed a LW escape fraction of unity throughout this 
paper. We do not expect this to have a strong effect on our results, but note that there could be an impact for cases with very low
star formation efficiency \citep{2015MNRAS.454.2441S}.

Finally, we note that we do not include subhalos in our model. The only halos considered are distinct halos 
which are not substructure of more massive halos. We also note that we do not allow ``splash-back'' halos 
(i.e. those subhalos which enter and then leave a larger halo) to contribute to star formation.
This is to prevent double-counting star formation. A splash-back halo contributes 
to star formation when it first enters a halo, and we do not include more star formation if it exits and re-enters.

\subsection{LW Feedback}
We calculate the LW flux incident on a minihalo as $J_{\rm LW, 21} = J_{\rm loc}(\vec{x}, z) + J_{\rm bg}(z)$, where $J_{\rm loc}(\vec{x}, z)$ is the local flux from nearby individual sources and $J_{\rm bg}(z)$ is the mean
 LW background on large scales generated by sources outside of the simulation box.
The uniform background from sources beyond the simulation box is given by 
\begin{equation}
\label{bg_eqn}
J_{\rm bg}(z) = \frac{c (1+z)^3}{4 \pi} \int_{z_{\rm Rmax}}^\infty dz' \epsilon_{\rm LW}(z') \left | \frac{dt_{\rm H}}{dz'} \right | f_{\rm LW}(z', z),
\end{equation}
where $\epsilon_{\rm LW}(z')$ is the mean LW emissivity in our box as a function of redshift, $t_{\rm H}$ is the Hubble time, and $f_{\rm LW}(z',z)$ the attenuation of LW flux observed at redshift $z$ from sources at redshift $z'$ due to LW photons being redshifted into Lyman series resonance lines and absorbed \citep{1997ApJ...476..458H}. We approximate this attenuation with Eq. 22 in \cite{2009ApJ...695.1430A}. The limit of integration, $z_{\rm Rmax}$, corresponds to the redshift of an object at a distance $R_{\rm max} = \sqrt{2}/2 L_{\rm box}$ from an observer at $z$, due to the finite light travel time. Thus, for each halo, the LW contribution from distances greater than $R_{\rm max}$ is given by $J_{\rm bg}$ and by $J_{\rm loc}$ for distances smaller than $R_{\rm max}$. 
When computing the integral in Eq. \ref {bg_eqn}, we smooth $\epsilon_{\rm LW}$ over a redshift range corresponding to a change in cosmic time of 10 Myr. This is meant to 
reduce rapid fluctuations, which are not expected in the background component (which has contributions from $\sim 100$ Mpc scales). 
Additionally, for very high redshift, when there are few sources, we set the large-scale background by hand to mimic the contribution 
 from stars formed outside of the simulation box. We find that setting $J_{\rm bg, 21}(z) = 850 \times 10^{-z/5.5} $ at $z>25$ maps smoothly onto the background
 flux calculated with Eq. \ref{bg_eqn} at lower redshifts in our fiducial model. This ad hoc procedure only impacts the very highest redshift in our simulations. 
 As discussed in Section 5, due to the small size of our box we cannot hope to 
 accurately determine the global LW background. In future work, we intend to  combine the semi-analytic 
 models described in this paper with larger-scale models ($>{\rm 100}$ Mpc) to simultaneously capture large-scale and small-scale astrophysical effects.

When computing the local LW contribution, $J_{\rm loc}(\vec{x}, z)$, we adopt a grid-based method. We break our simulation box 
into a cubic grid with $256^3$ cells. We find that the results have converged for this resolution, reducing the resolution to $128^3$ 
changes the Pop III and metal-enriched SFRD by $\sim10$ percent or less. This is expected since the $256^3$ grid has a spatial 
resolution of $12$ kpc (comoving), which is approximately equal to the virial radius of an atomic cooling halo at $z\approx 10$.
 For a pristine halo in cell $j$, 
the LW intensity coming from all other cells is well approximated as
\begin{equation} 
\label{J_eqn}
J_{j} = \sum_{i \ne j}   \frac{E_i}{ | \vec{r}_j - \vec{r}_i | ^2}, 
\end{equation}
where $E_i$ is the LW energy currently emitted from stars in cell $i$, per time, per frequency, divided by $(4\pi)^2$ and
$\vec{r}_i$ is the position of the $i$'th cell.
We note that using the convolution theorem, this can be rapidly computed using FFTs. 
We compute Eq. \ref{J_eqn} using
\begin{equation}
J_{j} =  (E *  |\vec{r} |^{-2})_j = \mathcal{F}^{-1} \left \{   \mathcal{F} \{E \}  \mathcal{F} \{ | \vec{r} | ^{-2} \} \right \}_j,
\end{equation}
where $*$ signifies convolution and $\mathcal{F}$ and $\mathcal{F}^{-1}$ denote the discrete Fourier and inverse Fourier transforms, respectively.
We compute these with FFTs on our $256^3$ grid, utilizing periodic boundary conditions to determine 
$|\vec{r}|^{-2}$ for each cell and setting the contribution from halos to their own cell (i.e., $|\vec{r}| = 0$) to zero.
We then add the contribution from halos within the same cell using the exact distance to ensure that close halo pairs are treated accurately.
In our fiducial model, there are very few minihalos with star-forming halos in the same cell (there are $\sim 100$ sources spread among 256$^3$ cells at $z\sim6$).
We find that this approximate method quickly computes essentially the 
same LW fluxes as would be obtained by a brute force calculation computing the distance squared between all relevant pairs of halos (for most halos
the difference in the computed local LW intensity between the two methods is less than 10 percent).

\subsection{IGM Ionization State}
As discussed above, regions of the IGM that have been reionized will be photoheated, 
which can suppress star formation in small dark matter halos ($<M_{\rm ion}$ in our model parameterization).
In order to compute the ionization fraction throughout our box, we utilize a grid-based method 
inspired by larger-scale semi-numerical approaches used in the past \citep[e.g.,][]{2004ApJ...613....1F,2005ApJ...630..657Z, 2007ApJ...669..663M, 2011MNRAS.411..955M}.
At each redshift, we begin with a $256^3$ resolution cubic grid $P$, where each cell contains the 
total number density of ionized photons previously produced in all the dark matter halos within that cell that
have escaped into the IGM. This is given by $P_{\rm i} = f_{\rm esc, II}\eta_{\rm II}M_{\rm *, II, i}/(m_{\rm proton}V_{\rm cell}) + f_{\rm esc, III}\eta_{\rm III}M_{\rm *, III, i}/(m_{\rm proton}V_{\rm cell})$, where $M_{\rm *, II, i}$ and $M_{\rm *, III, i}$ are the total masses of metal-enriched and Pop III stars ever formed in cell i and $V_{\rm cell}$ is the comoving density of each cell. Here $f_{\rm esc, III}$ and $f_{\rm esc, II}$ denote the escape fraction of hydrogen ionizing photons from halos hosting Pop III and metal-enriched stars, respectively.
We smooth $P$ on a range of scales, corresponding to different ionized bubble sizes, $R_{\rm bub,i}$.  
After smoothing on a given bubble size, we identify cells which are the center of $R_{\rm bub,i}$-sized 
spheres which have a total number of ionizing photons per bubble volume 
greater than a threshold $i_{\rm thresh}$. 
These cells are set to be the centers of ionized bubbles of radius $R_{\rm bub,i}$. 
If one follows this procedure for an isolated source of ionizing radiation, selecting $i_{\rm thresh}= \bar{n}_{H}/8$ (where $\bar{n}_{\rm H}$ is the cosmic mean comoving density of hydrogen atoms) gives an ionized bubble around 
the source with a size that exactly corresponds to the number of ionized photons produced by the source (ignoring recombinations). 
However, if there are a number of sources which are distributed over a spatially extended region that contribute to the same ionized bubble,
one must select a higher value of $i_{\rm thresh}$ to correctly recover the bubble size. 
Thus, the exact value of $i_{\rm thresh}$ which should be used depends on the clustering of the sources within the box. 
We find that using $i_{\rm thresh}= \bar{n}_{H}/4$ leads the total ionization fraction to be within $\approx 20 \%$ of the total 
number of escaped ionizing photons in the box divided by the total number of hydrogen atoms in the box. 

We find that our model converges for 30 logarithmically-spaced bubble radii ranging 
from one cell to half the length of the box. The smoothing mentioned above is rapidly computed with FFTs 
by applying the convolution theorem to $P$ and a spherical top-hat window function centered around the origin. 
Note that this method does not include recombinations. 
In \cite{2018MNRAS.475.5246V}, we found that recombinations did not have a significant impact on our semi-analytic results for the redshifts explored ($z >20$). 
This is due to the rapid increase in star formation as a function of time. 
By the time a region could recombine, a much larger number of ionizing 
photons have been produced in the same region.
At later times adding recombinations could potentially slow the growth of bubbles in our simulations, however 
given the other uncertainties in the modeling (e.g., the time evolution and halo mass dependance of the escape fractions of ionized photons), we 
defer a detailed treatment to future work.

\subsection{IGM Metallicity}
Our semi-analytic model permits pristine halos to be externally polluted by SN winds, leading to metal-enriched star formation.
Thus, it is necessary to compute the metallicity of the IGM throughout the simulation box.
To accomplish this, we assume that SN winds create spherical metal bubbles around halos
which begin expanding 4 Myr after star formation begins (this delay is caused by the lifetime of massive stars).
Following \cite{2009ApJ...700.1672T}, we assume that the bubble velocity is given by $v_{\rm bub} = f_{\rm bub}60~{\rm km~s^{-1}}$ until it reaches a comoving radius of $R_{\rm bub} = f_{\rm bub}150~h^{-1}~{\rm kpc}$. We include a free parameter $f_{\rm bub}$, which we vary to determine the impact of different bubble sizes on our results. We find that the bubble radius given by this procedure (with $f_{\rm bub}=1$) is in rough agreement with the hydrodynamic cosmological simulations of \cite{2015MNRAS.452.2822S}, which include radiative transfer and resolve the blast-wave of the SN explosion. 
To compute the total amount of metals in each bubble, we assume $10~M_\odot$ of heavy elements are produced per each $40~M_\odot$ (chosen as the Pop III stellar mass for simplicity) of Pop III stars \citep{2006NuPhA.777..424N}  and $1~M_\odot$ of metals are produced for every $100~M_\odot$ of Pop II stars \citep[this is the approximate metal yield assuming that stars above $8~M_\odot$ lead to a supernova and $\sim 1~M_\odot$ of heavy metals are produced per supernova][]{2001PhR...349..125B}. We assume metals are spread uniformly throughout each bubble and the metallicity is summed where bubbles overlap.

\begin{figure}
\centering
\includegraphics[clip=false,keepaspectratio=true, width =  60mm]{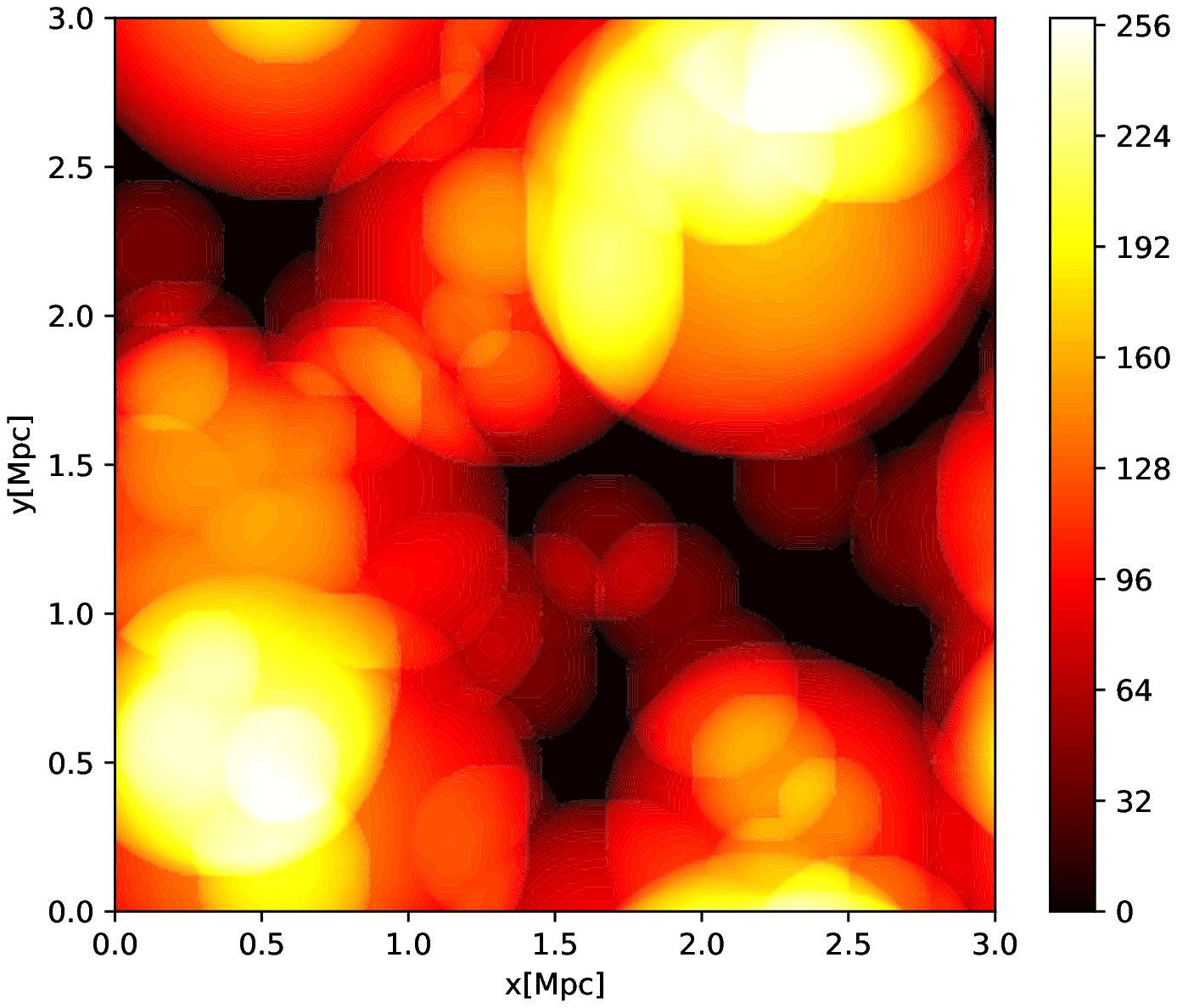}
\includegraphics[clip=false,keepaspectratio=true, width = 60mm]{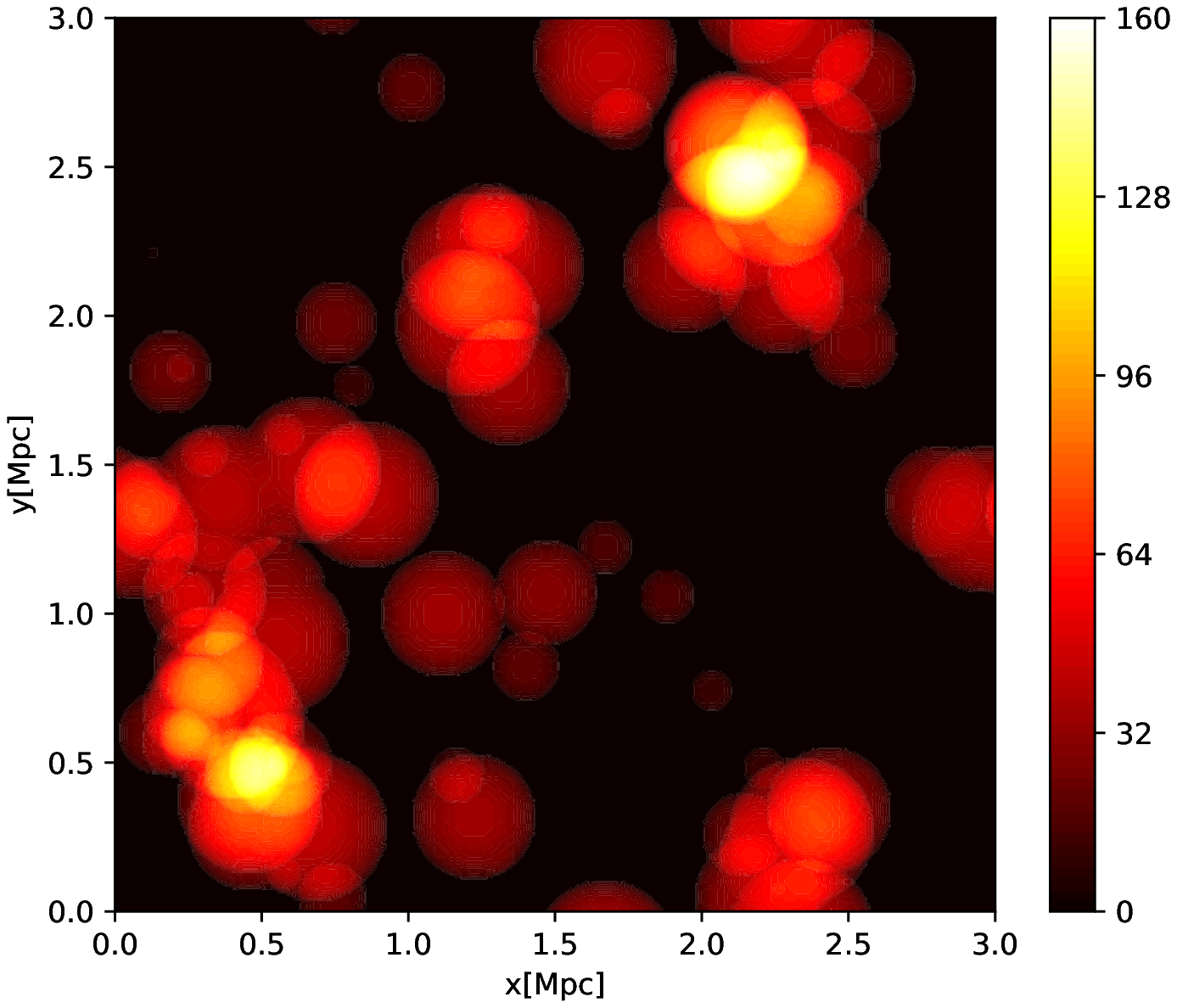}
\includegraphics[clip=false,keepaspectratio=true, width = 50mm]{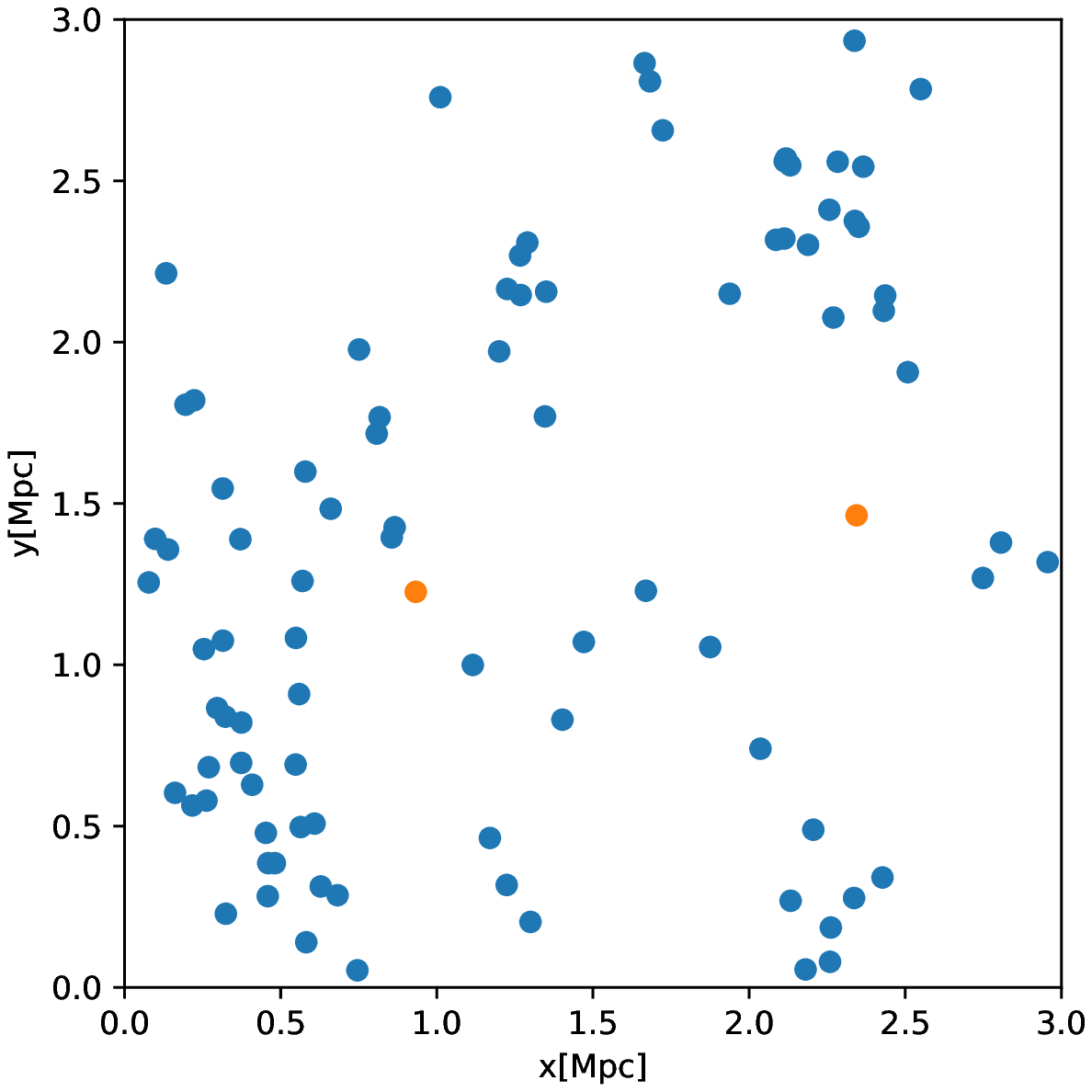}
\caption{\label{bubs} Projections of the ionized fraction of the IGM (left panel), regions of the IGM enriched above $Z = 10^{-6}~ Z_\odot$ (middle panel) and locations of Pop III (orange points) and metal-enriched stars (blue points) formed in a single snapshot at $z = 8.6$ with fiducial model parameters (right panel). 
The value of the color bars correspond to the number of grid cells which are ionized or metal-enriched along the projected direction (with 256 as a maximum value set by the grid resolution of $256^3$).
For the fiducial model, ionized regions of the IGM generally extend beyond those that are enriched to $Z > Z_{\rm crit}$.
}
\end{figure}

Similar to our approach to LW and ionization feedback, we utilize a cubic grid of $256^3$ resolution elements, and calculate the metallicity 
in each cell according to the metal bubble properties just described.
We rapidly compute this grid by going though the same range of bubble sizes used for reionization, which are described in the preceding subsection. 
For each bubble size we take a grid populated with the centers of bubbles in that radius bin and smooth using the convolution 
theorem and FFTs. In Figure \ref{bubs}, we show projections of the ionization fraction and metallicity of the IGM computed 
with the grid-based methods discussed.

\subsection{Computational Cost}
Next we discuss the computational cost and scaling of our implementation of the method just described. The FFT-based method has a more efficient scaling with the size of the simulation compared 
to a simple approach which directly computes distances between pairs of dark matter halos.
For instance, the naive approach to computing the LW fluxes on potential-Pop III-forming halos requires computing
the distances between  $\approx N_{\rm sources}\times N_{\rm minihalos}$ pairs of halos. So for fixed
mass resolution, the compute time for this approach scales with the volume of the simulation box as $t_{\rm cpu} \propto V_{\rm box}^2$ (metal enrichment or reionization have the same scaling).
For our FFT-based method however, populating the grid with sources of LW/ionizing radiation or metal bubbles scales as $t_{\rm cpu} \propto V_{\rm box}$
and the compute time of FFTs scales as $\propto N \log (N)$, where $N$ is the number of elements in the grid. Thus, for fixed resolution our
FFT based approach scales as least as fast as $t_{\rm cpu} \propto V_{\rm box} \log (V_{\rm box})$. We note that the shell method used in \cite{2018MNRAS.473.5308M} also performs better than the simple $\propto N_{\rm sources}\times N_{\rm minihalos}$ case for both external metal enrichment and reionization. This is possible because one can ignore pairs of halos which are at significantly larger distances than the relevant bubble sizes.

We note that, due to our small box, this improved scaling was not particularly important in our fiducial model. It takes $\sim 1$ second to compute the distance between
$\sim 10^8$ pairs of halos on one of the cores of the Flatiron Institute's computer cluster ``Rusty''. It takes roughly the same time to compute an FFT on the $256^3$ grid used (note that the grid resolution does not need to match the N-body simulation's resolution). 
By the end of our simulation ($z\approx6$) in the fiducial parameterization, we have roughly $\sim100$ sources and $\sim10^5$ minihalos which must be checked for LW flux/metal enrichment (in each of our 10 realizations). 
For this small number of pairs, directly computing the distances is fast. 
However the improved scaling leads to a significant speedup in models where many halos are externally metal enriched, leading to many more sources.
In the cases presented below where metal bubbles reach minihalos before ionizing radiation 
we find many more halos with star formation (on the order of $\sim 10^4$ sources by $z \sim 6$).
In this case, where $\sim 10^9$ pairs of distances must be determined, the FFT approach is roughly an order of magnitude faster. In either case, 
for the box sizes we have chosen,
the compute times are relatively modest. 
Thus, the fact that the FFT method can more accurately capture inhomogeneous reionization compared to previous shell models is more 
important for the results presented below.
However, we point out that if much larger boxes were used, 
the scaling improvement of our FFT method would be much more pronounced.

\section{Results}
\begin{figure}
\plotone{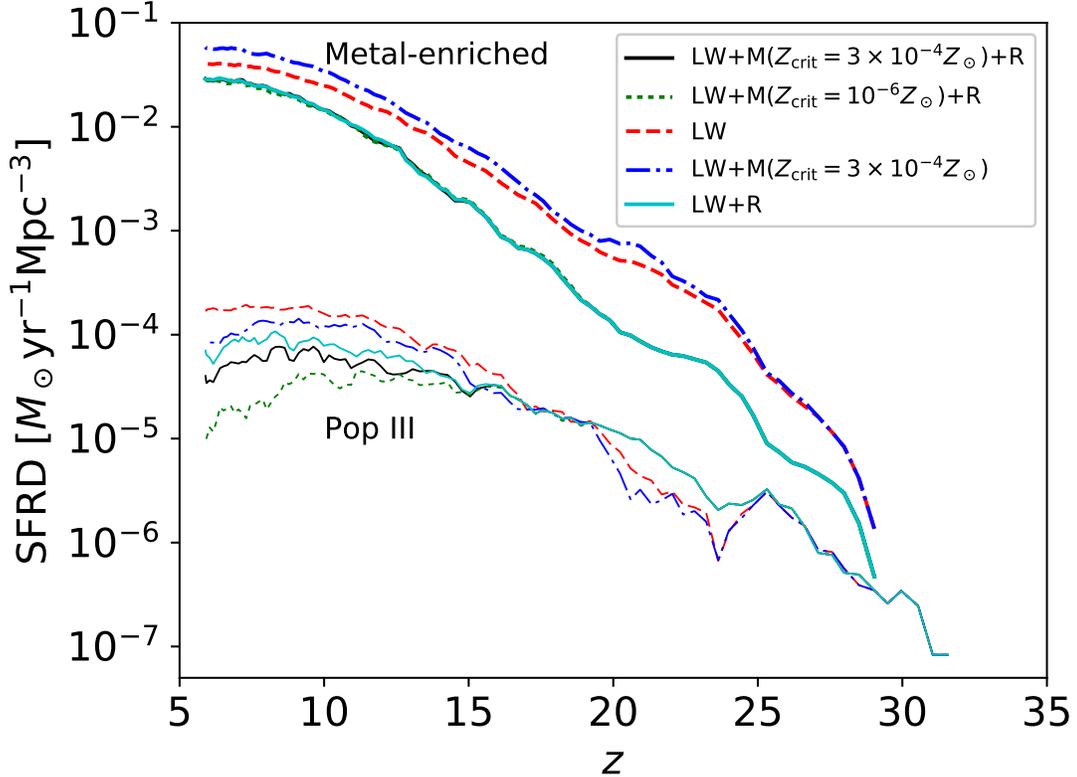}
\caption{\label{feedback} Cosmic SFRD for Pop III (thin curves) and metal-enriched stars (thick curves) including/excluding different combinations of LW feedback, external metal enrichment (M) and reionization feedback (R). These curves give the average quantities for ten realizations (different random seeds in the N-body simulations) of a 3 Mpc patch of the Universe. The curves have been smoothed with a boxcar average over a duration of $0.1 t_{\rm H}(z)$, corresponding roughly to a dynamical time for the mean density of a dark matter halo at the time of virialization.
}
\end{figure}

We begin by exploring how the cosmic star formation history is affected by various feedback
mechanisms. In Figure~\ref{feedback}, we show the evolving SFRD when including/excluding different combinations 
of LW feedback, external metal enrichment via supernovae winds, and cosmic reionization (while otherwise keeping the fiducial parameters in Table \ref{table}). 
The qualitative effects of these mechanisms can be understood as follows.
Compared to the LW only case, including external metal enrichment slightly increases the metal-enriched SFRD and reduces the Pop III SFRD due to small halos neighboring star-forming halos being externally enriched. 
When only including LW and reionization feedback, the metal-enriched SFRD is significantly reduced compared to LW alone. This is due to the quenching of star formation in small halos due to gas photoheating described in section 3.1. We note that this reionization feedback is mainly local at high redshift ($z\gtrsim15$), halos ionize their surrounding regions of the IGM leading to the subsequent quenching of star formation.
Compared to LW alone, there is also a small increase in the Pop III SFRD at $z\approx22$, which results from the decreased LW background intensity at this redshift. This ultimately leads to a lower Pop III SFRD at $z<15$ because more of the halos have already been metal-enriched at higher redshift.

When adding all three feedback processes, the metal-enriched star formation history does not change much from the LW and reionization case. This is because the metal bubbles do not extend as far as the ionization bubbles in our fiducial model (see Figure \ref{bubs}). Thus, small halos neighboring star formation which are enriched by metals are also ionized and do not form any stars. Combining reionization and metal enrichment reduces the Pop III SFRD even more than each alone at $z \leq 15$. In this case, the ionization and external metal enrichment are acting in tandem. The ionization feedback first reaches a pristine halo and delays its Pop III star formation. Metals spread through the IGM arrive before the halo has reached a sufficient mass to form stars.

We note that both the metal-enriched and Pop III SFRDs have a feature at $z \sim 22-25$. This is due to the LW background put in by hand and changing at $z=25$ as described above. Putting this in by hand was necessary due to the small size of our box, however in future work we plan to combine these models with techniques incorporating larger spatial scales to address this issue.

The Pop III and metal enriched SFRDs in Figure \ref{feedback} can be compared to a number of previous works.
Observationally, \cite{2016PASA...33...37F} reports the total SFRD density from the integration of UV luminosity functions (LFs) down to ${\rm M_{UV}=-13}$ \citep[see also][]{2014ARA&A..52..415M}. These observational values decline from $\sim 10^{-1.5} M_\odot {\rm yr^{-1} Mpc^{-3}}$ at $z=6$ to $\sim 10^{-1.9} M_\odot {\rm yr^{-1} Mpc^{-3}}$ at $z=10$. This is very similar to our fiducial model with all feedback mechanisms included (metal-enriched SFRD of $\sim 10^{-1.6} M_\odot {\rm yr^{-1} Mpc^{-3}}$ and $\sim 10^{-1.8} M_\odot {\rm yr^{-1} Mpc^{-3}}$ at $z=6$ and $z=10$, respectively).
While currently there are no observations of the Pop III SFRD, we can compare our results to previous theoretical models. We find that in our fiducial model, the Pop III SFRD is quite similar to the ``classical" Pop III model from the hydrodynamical cosmological simulations of \cite{2019ApJ...871..206S} that do not include their sub-grid prescription for inefficient mixing of metal pollutants. Our results agree mostly within a factor of a few for most redshifts between $z=6-20$ (see their Figure 1). We note that the agreement in metal enriched SFRD is not as close (significantly higher SFRD at $z\gtrsim10$ in our model). A qualitatively closer match could be obtained by increasing our $t_{\rm delay}$ (as discussed below). We also note good agreement with the Pop III SFRD from the numerical simulations of \cite{2013ApJ...773...83X}, matching to roughly a factor of a few over $z=15-30$ (see their Figure 2). In contrast to \cite{2019ApJ...871..206S}, \cite{2013ApJ...773...83X} find higher metal-enriched SFRD than our fiducial model. When comparing to the analytic model of \cite{2003ApJ...589...35S}, compensating for a factor of 100 difference in star formation efficiency, we find a similar Pop III SFRD to their model with the lowest energy driving metal bubbles into the IGM. A precise comparison is challenging however, due to major differences in modeling feedback processes (e.g, the impact of reionization). Broadly speaking there is qualitative agreement between our results and a number of other simulations and semi-analytic models \citep[e.g., ][]{2013MNRAS.428.1857J, 2014MNRAS.440.2498P, 2018MNRAS.475.4396J, 2018MNRAS.473.5308M}, but there remain order of magnitude quantitative variations in the high-redshift Pop III and metal enriched SFRDs \citep[see Figure 1 in][]{2019ApJ...871..206S}. This is not surprising due to the uncertainty of the properties of the first stars and illustrates the importance of rapid semi-analytic models to efficiently survey the relevant model parameter space.

\begin{figure*}
\centering
\includegraphics[clip=false,keepaspectratio=true, width = 195mm]{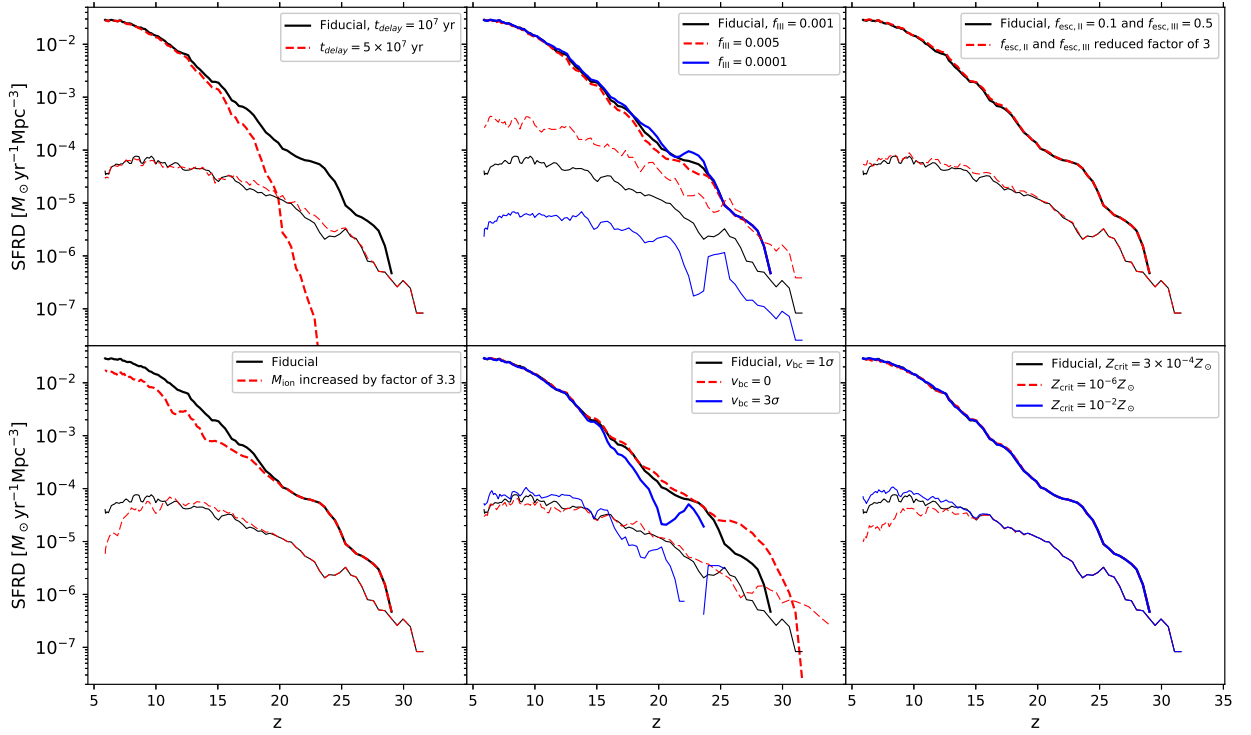}
\caption{\label{variations} Cosmic SFRD for parameter changes around the fiducial model (top curves metal-enriched star formation, bottom curves Pop III). 
All feedback mechanisms (LW + external metal enrichment + reionization feedback) are included. The averaging across realizations and smoothing over time is the same as Figure \ref{feedback}. The parameters varied in each panel are (clockwise starting in the top-left panel) $t_{\rm delay}$, $f_{\rm III}$, $f_{\rm esc,II}/f_{\rm esc,III}$, $M_{\rm ion}$, $v_{\rm bc}$, and $Z_{\rm crit}$.   }
\end{figure*}

Next, we examine how the star formation history is impacted for parameter variations (mostly changing one parameter at a time) around our fiducial model (with all feedback mechanisms included).
These results are shown in Figure \ref{variations}, where a number of parameters are adjusted. 
As expected, the Pop III star formation efficiency, $f_{\rm III}$, is the most important parameter impacting the amount of Pop III stars formed over cosmic time.
We note that many of the parameters (i.e.~$f_{\rm esc,II}$, $f_{\rm esc,III}$,  $t_{\rm del}$) do not have a strong impact on the cosmic abundance of Pop III stars in our fiducial model. The main exceptions are that at high redshift (above $z\approx 20$) a high-streaming velocity, $v_{\rm bc}$,  greatly reduces the abundance of Pop III stars. At lower redshift ($\lesssim 12$), the abundance of Pop III stars depends on $Z_{\rm crit}$ and $M_{\rm ion}$. This is because either reducing $Z_{\rm crit}$ or increasing $M_{\rm ion}$ increases the number of pristine halos which are externally metal-enriched (in the latter case because Pop III star formation is delayed allowing metals to arrive). We note that when varying $M_{\rm min, met}$ within the fiducial model for the values indicated in Table \ref{table}, there are minimal changes to either the Pop III or metal-enriched SFRD.

In the fiducial model, metal-enriched star formation quickly dominates over Pop III. However, we point out that the global transition from Pop III to metal-enriched star formation seems to be strongly controlled by the delay between Pop III and metal-enriched star formation due to SN feedback, $t_{\rm delay}$. Increasing this timescale by a factor of five drastically changes metal-enriched star formation at early times. This is partly due to our specific prescription for reionization feedback. Pop III star formation generally reionizes the surrounding region and if $t_{\rm delay}$ is increased, the star formation suppression factor due to photoheating ($\exp (-t_{\rm rei}/[0.1t_{\rm H}])$) becomes much smaller by the time metal enriched star formation can occur.

\begin{figure*}
\centering
\includegraphics[clip=false,keepaspectratio=true, width =  75mm]{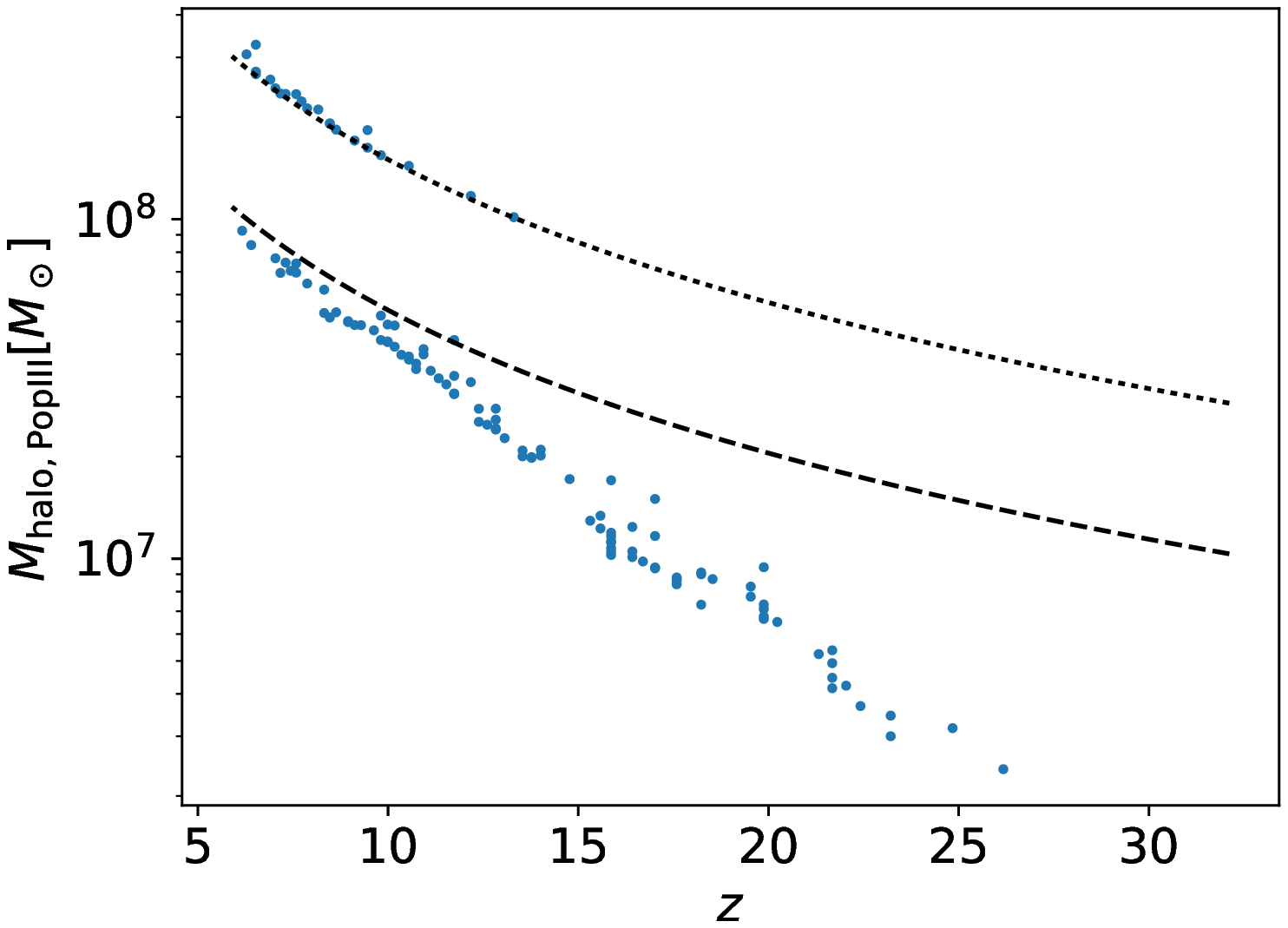}
\includegraphics[clip=false,keepaspectratio=true, width = 75mm]{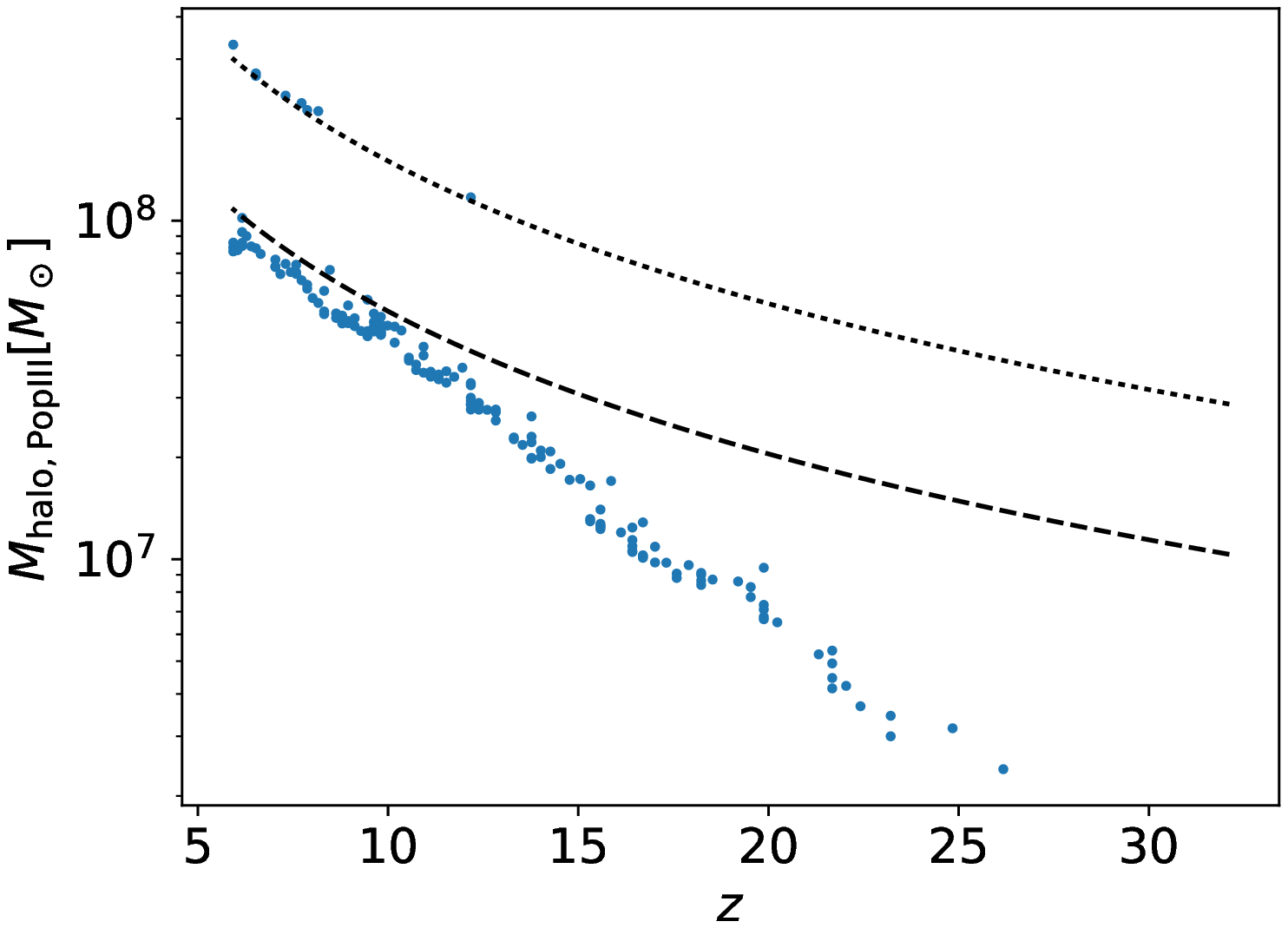}
\caption{\label{popIII_masses}  Dark matter halo masses and redshifts hosting Pop III star formation from one of our 10 simulation realizations. We show the fiducial model (left panel) and a model where $f_{\rm esc, II}$ and 
$f_{\rm esc, III}$ are reduced by a factor of three (right panel). The dashed and dotted curves show $M_{\rm a}$ and $M_{\rm ion}$, respectively. Even though less halos host Pop III stars formation at $z \lesssim 10$ due to reionization feedback in the fiducial model, the total SFRD is similar because the star formation occurs in larger halos.
}
\end{figure*}

Two parameters that warrant further discussion are the escape fractions of ionizing radiation, $f_{\rm esc, II}$ and $f_{\rm esc, III}$. From Figure \ref{variations}, we see that the total amount of star formation is not strongly impacted by varying this parameter by a factor of three. Since the ionized bubbles envelop the entire box by $z=6$ this result initially seems unexpected. However, closer inspection reveals that while the total abundance of Pop III stars does not change, the halos where they form does. Thus, in the case with a lower escape fraction more Pop III stars form in lower mass halos, but because there is less gas mass, a lower amount of Pop III stars forms in each of these halos. The reason the abundance is so similar is due to the shape of the halo mass function. Over the relevant redshift and mass range, $M\frac{dn}{dM} \approx {\rm constant}$. Thus, if the minimum halo mass for star formation increases by some factor, the reduction in the amount of halos forming Pop III stars is very closely compensated by the increased amount of star formation due to the larger halo masses. This can be seen in Figure \ref{popIII_masses}, where we show masses and redshifts of halos hosting Pop III star formation in the fiducial model compared to reducing the ionizing escape fractions by a factor of 3. At $z\lesssim 10$ the fiducial model has fewer Pop III-forming halos, but they have closely compensating higher masses leading to a very similar Pop III SFRD (see Figure \ref{variations}).

\begin{figure*}
\centering
\includegraphics[clip=false,keepaspectratio=true, width =  75mm]{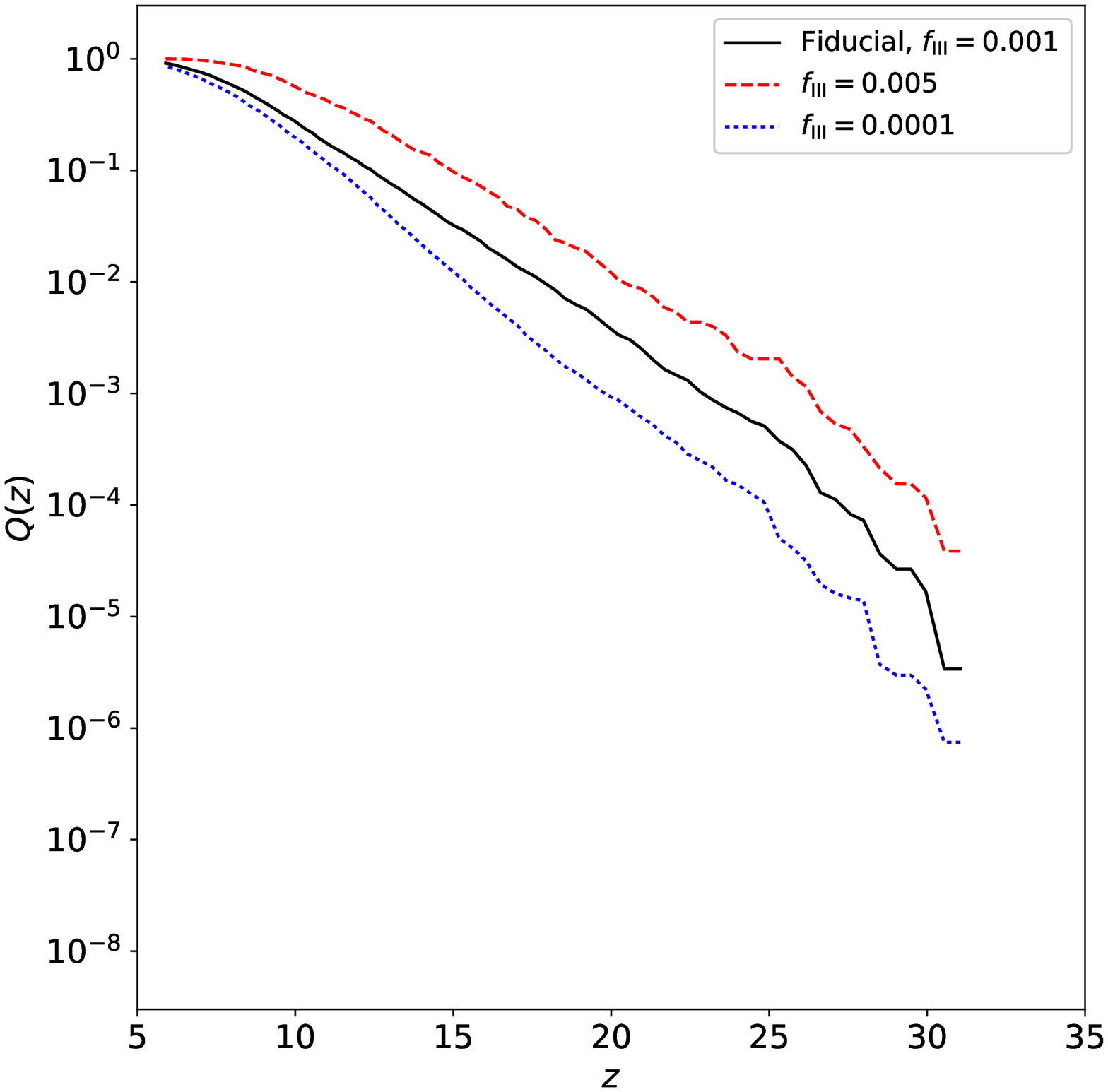}
\includegraphics[clip=false,keepaspectratio=true, width = 75mm]{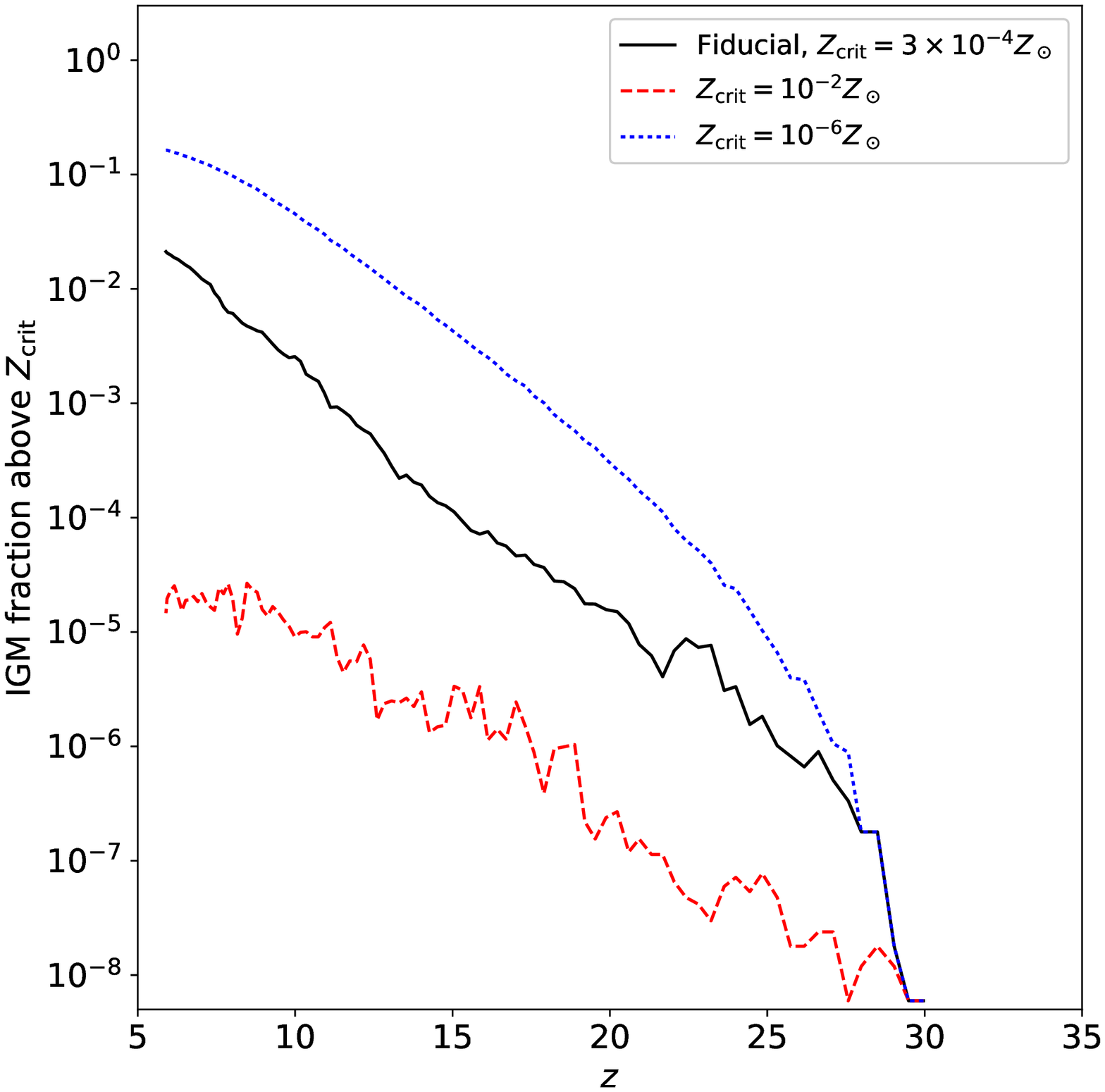}
\caption{\label{ion_avg} Mean fraction of the volume of the IGM that is ionized (left panel) and metal-enriched above $Z_{\rm crit}$ (right panel).
}
\end{figure*}

We also show how the fraction of the simulation box which is ionized and metal-enriched evolves over time (see Figure \ref{ion_avg}). As mentioned above, in our models the ionization fraction is generally much larger than the fraction of the IGM which is significantly enriched by metals. Even though the simulation box is far too small to accurately characterize reionization, we see that the box is completely ionized at $z\approx6$ \citep[which is roughly consistent with observational constraints such as the optical depth from][]{2016A&A...596A.108P}.

\begin{figure*}
\centering
\includegraphics[clip=false,keepaspectratio=true, width =  75mm]{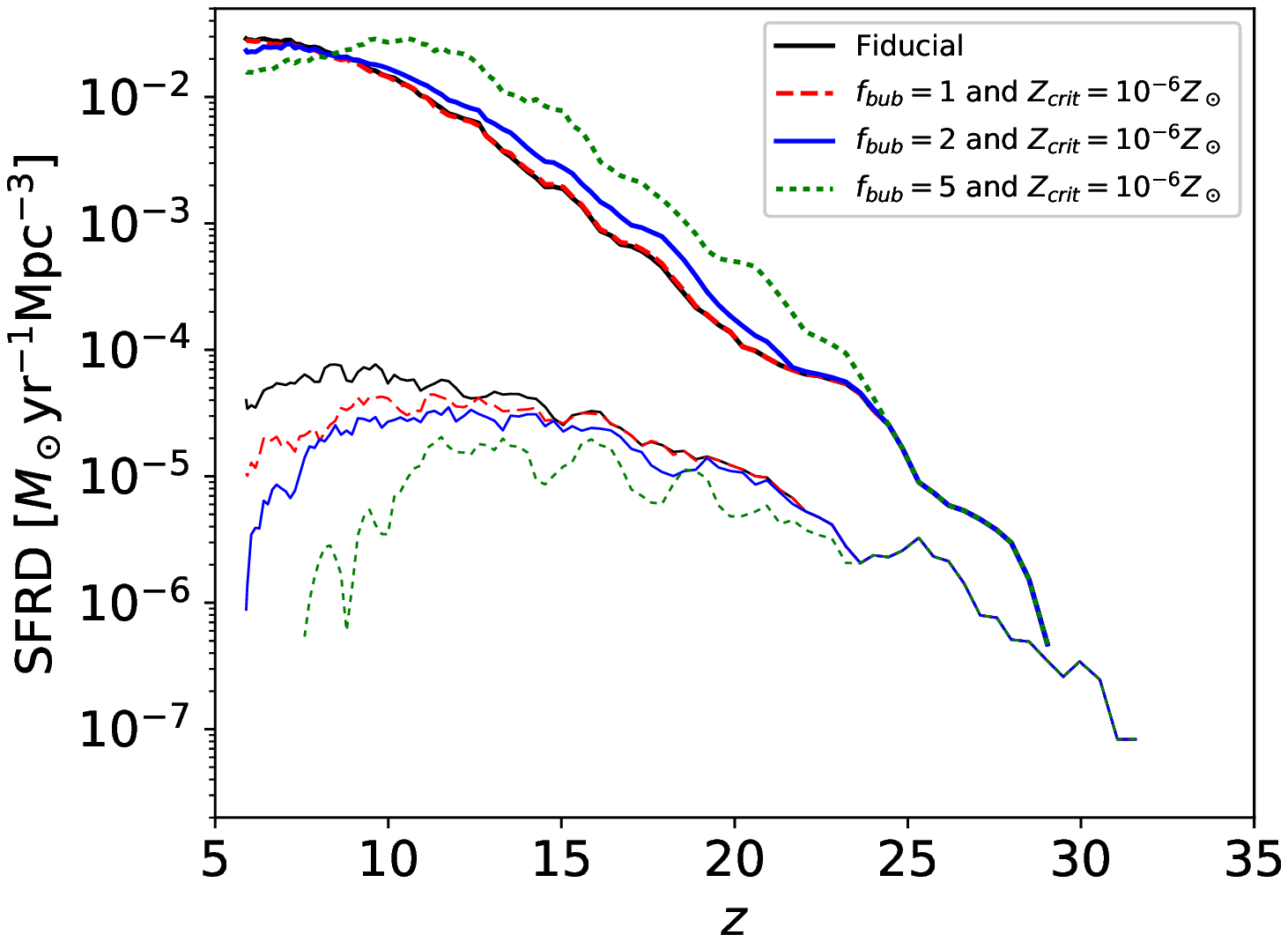}
\includegraphics[clip=false,keepaspectratio=true, width = 75mm]{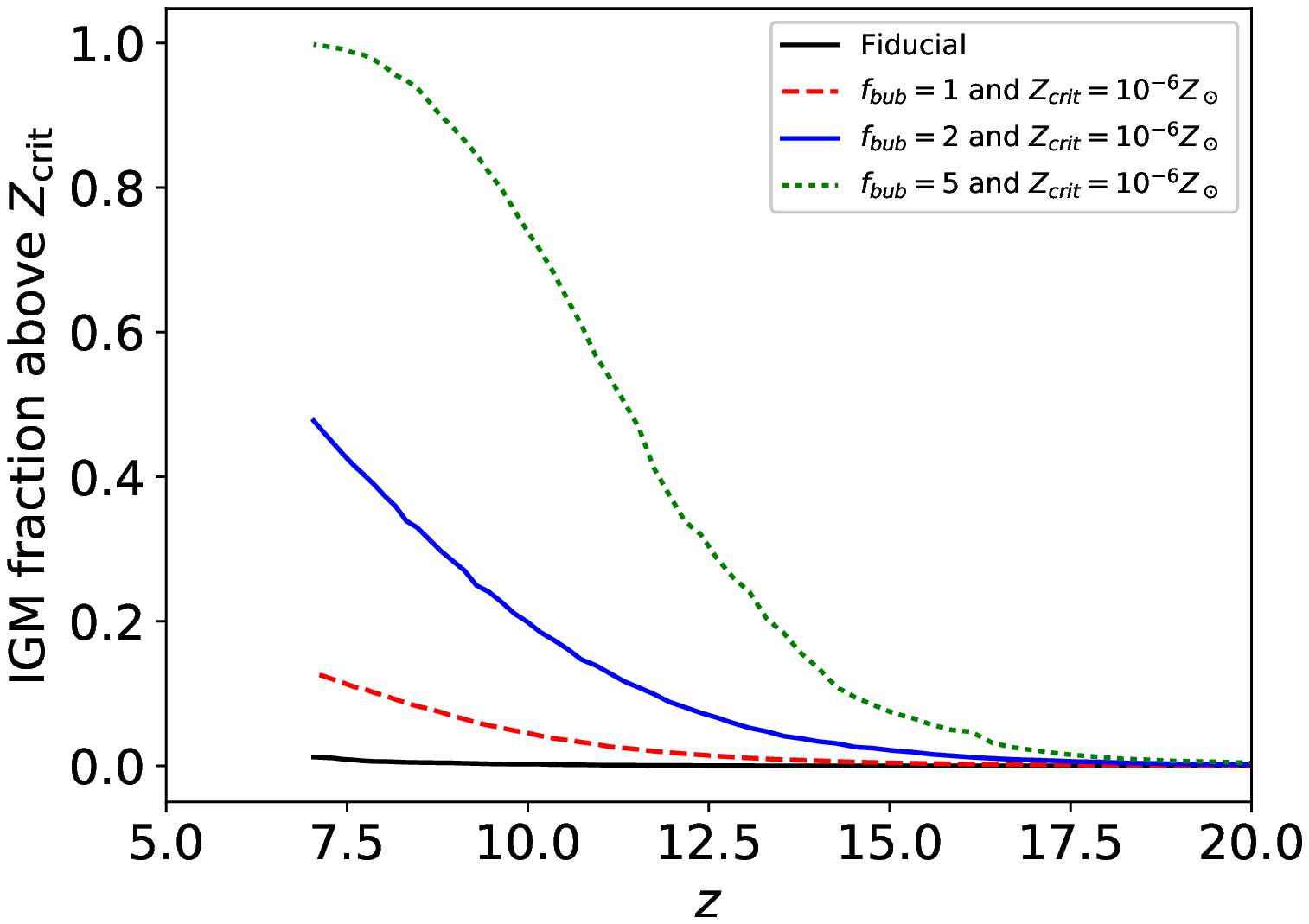}
\caption{\label{more_bubs}  \emph{Left panel:} The Pop III (thin lines) and metal-enriched (thick lines) SFRD when varying $f_{\rm bub}$ and $Z_{\rm crit}$ simultaneously. It is clear that larger metal bubbles driven by SN winds can greatly reduce the abundance of Pop III star formation if the critical metallicity is the value expected for dust cooling. \emph{Right panel:} The volume fraction of the IGM enriched above $Z_{\rm crit}$. Note that for $f_{\rm bub}  = 5$ and $Z_{\rm crit}=10^{-6}~Z_\odot$ the entire box is enriched by metals by $z\approx 7.5$, quenching Pop III star formation. 
}
\end{figure*}

We have also considered cases varying $f_{\rm bub}$ and $Z_{\rm crit}$ simultaneously. Increasing $f_{\rm bub}$ alone 
has very little impact on the Pop III SFRD because, for the fiducial $Z_{\rm crit}$, as the bubbles grow they dilute below the critical metallicity 
reducing the impact for external metal enrichment. This is not the case when simultaneously reducing the $Z_{\rm crit}$  to the level 
corresponding to dust cooling and increasing $f_{\rm bub}$. This is illustrated in Figure \ref{more_bubs}, where we see that,
in extreme cases, external metal enrichment can quench Pop III star formation at earlier times or lead to an increase
in early metal-enriched star formation due to external enrichment of minihalos. The latter effect only occurs if the metal bubbles 
expand faster than ionization bubbles to prevent the suppression of star formation discussed above.

In Figure \ref{LF} we include a comparison of our fiducial model to observations of the UV LF. We have converted SFR to UV magnitude via $SFR/(M_\odot {\rm yr}^{-1}) = 2.24\times10^{-18}~L_\nu/({\rm erg s^{-1} Hz^{-1}})$.
This relation is based on the ratio of the dust-corrected SFRD to the luminosity density given in \cite{2015ApJ...803...34B}.
We note that aside from using a star formation efficiency that was set through abundance matching, we have made no detailed efforts to match to observations. It is encouraging to see agreement without more specific tuning.  We do not dwell on this result as we have mainly focused on the physics of Pop III star formation in our modeling, but note that our simple model predicts a flattening for galaxies fainter than a rest-frame UV magnitude of $M_{\rm AB} \approx -10$.
	
\begin{figure}
 \epsscale{0.6}
\plotone{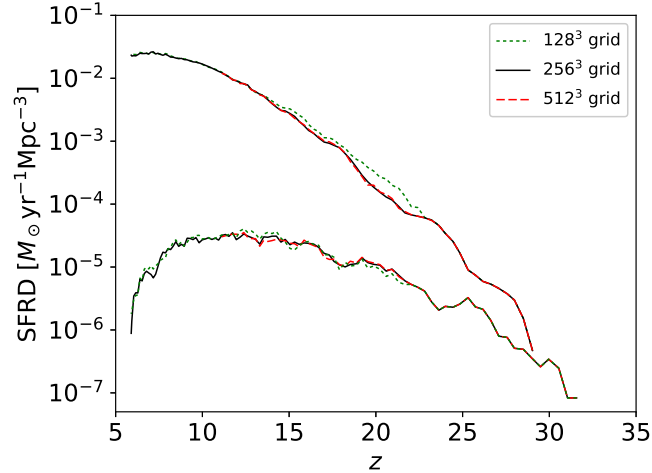}
\caption{\label{converge} The Pop III (lower set of curves)and metal-enriched (upper set of curves) SFRD for various resolutions in our grid-based prescriptions for LW feedback, reionization, and external metal enrichment. The model plotted has fiducial parameter values except for $f_{\rm bub}=2$ and $Z_{\rm crit} = 10^{-6}Z_\odot$. These values were selected to explore a model where metal bubbles in the IGM have a particularly large impact. We note excellent agreement between grid resolutions of $256^3$ and $512^3$, justifying our choice of $256^3$ in the other results presented in this paper.
}
\end{figure}

\begin{figure}
 \epsscale{0.6}
\plotone{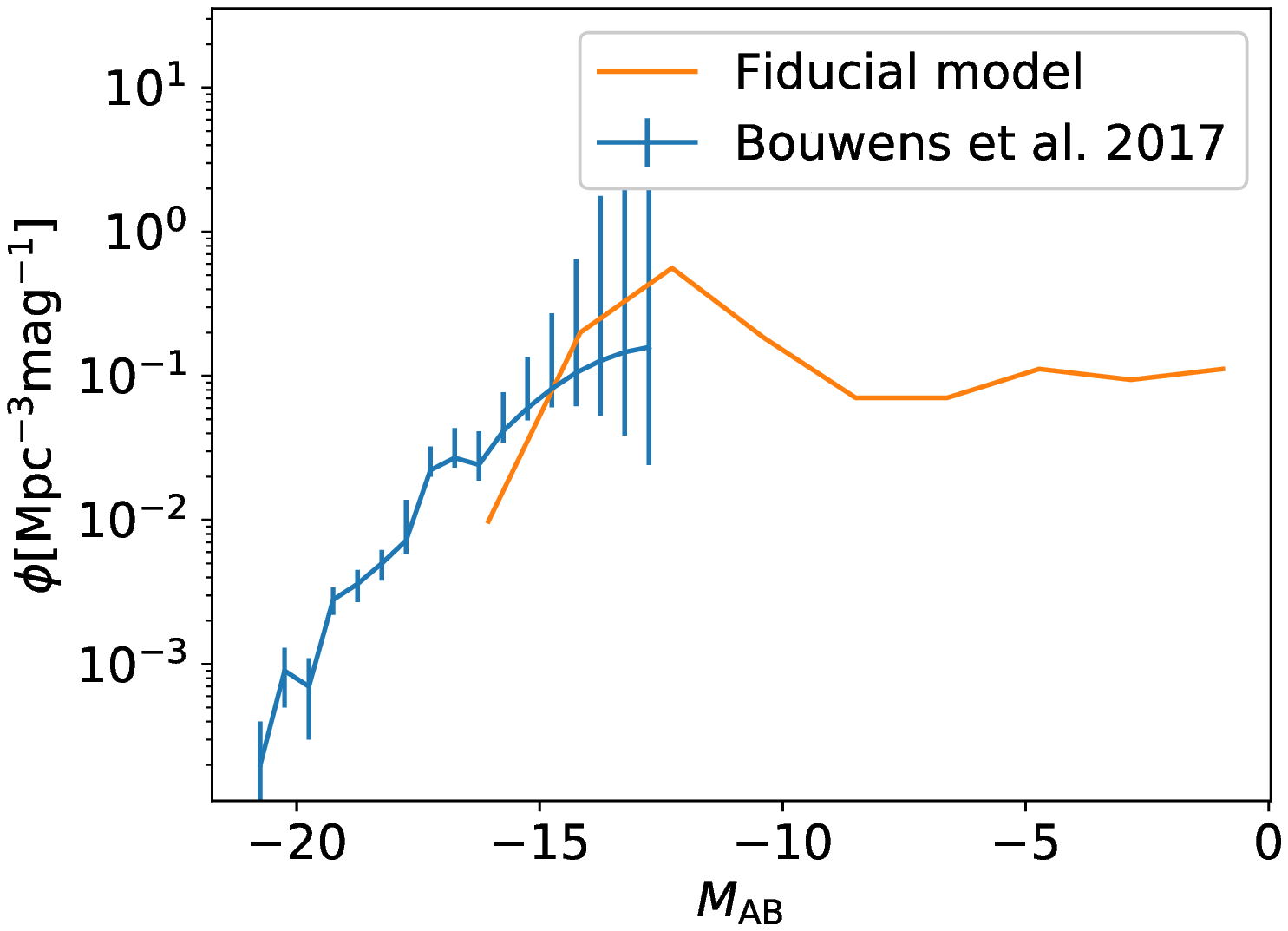}
\caption{\label{LF} Rest-frame UV luminosity function of metal-enriched stars from our fiducial model (all feedback processes included) compared to observational data at $z=6$ (with 1$\sigma$ errors) from \cite{2017ApJ...843..129B}.
}
\end{figure}

We conclude this section by examining our choice of resolution in the grid-based feedback prescription for LW feedback, reionization, and external metal enrichment. In Figure \ref{converge}, we plot the Pop III and metal enriched SFRDs for a model which differs from the fiducial case in two parameters ($f_{\rm bub}=2$ and $Z_{\rm crit} = 10^{-6}Z_\odot$) for various grid resolutions. External metal pollution is more extreme for these parameters compared to the fiducial model as discussed above. Thus, this parameterization provides a good check for convergence of external enrichment. Figure \ref{converge} shows good agreement for the resolutions plotted ($128^3$, $256^3$, and $512^3$) at most redshifts. However, we note that for the coarsest resolution, the metal-enriched star formation is not converged at $z\approx20$. We see very good agreement between the $512^3$ and  $256^3$ resolutions at all redshifts. The agreement is generally better than ten percent and justifies our choice of $256^3$ for the other results presented above.
	
\section{Discussion and Conclusions}
In this paper, we have presented a new semi-analytic model of Pop III and metal-enriched star formation in the early Universe.
This model takes in dark matter halo merger trees (including 3-dimensional clustering information)
obtained from cosmological N-body simulations and applies analytic models to populate the halos with 
stars. The main new feature of these simulations is a grid-based approach 
to compute feedback from LW radiation, photoheating of gas due to inhomogeneous reionization, and metal bubbles 
in the IGM due to SN winds. These are the first semi-analytic models of Pop III star formation with both external metal enrichment 
and a prescription for reionization which takes into account inhomogeneous reionization properly including multiple ionizing sources
contributing to the same HII region. We also note that the improved scaling of our grid-based technique ($t_{\rm cpu} \propto V \log V$ compared
to $t_{\rm cpu} \propto V^2$ in the naive case) permits larger boxes in 
future work.

We applied our semi-analytic model to N-body simulations which resolve $\approx 10^5~M_\odot$ dark matter minihalos
for a wide range of parameterizations spanning the parameters summarized in Table \ref{table}. 
We focus on results pertaining to Pop III stars, as 
our model is uniquely positioned to follow  the important feedback processes for metal-free 
star formation. Overall, we find that adding a full 3-dimensional treatment of reionization 
and external metal enrichment can have a strong impact on the cosmic star formation history. 
Reionization feedback
significantly reduces the metal-enriched SFRD at all redshifts. We note that at high redshift ($z\gtrsim 20$), this reduction does not depend on the clustering of halos
because it is caused by halos' direct progenitors (and thus reducing the ionizing photon escape fraction does not affect the Pop III or metal-enriched SFRD in Figure \ref{variations}). This is consistent with our results from \cite{2018MNRAS.475.5246V}.
At $z\lesssim 15$, external metal enrichment and 
reionization work in tandem to reduce the Pop III SFRD by roughly an order of magnitude compared to 
LW feedback alone. Reionization delays star formation in pristine halos which 
are then enriched by metal bubbles before any Pop III stars can form.

Consistent with \cite{2018MNRAS.473.5308M}, we both find that at late times there can
be sustained Pop III star formation in regions of the IGM which have been reionized. This motivates additional hydrodynamical simulations
of Pop III stars formed in these environments \citep[for simulations in the case of high ionizing flux see][]{2017MNRAS.469.1456V, 2019ApJ...882..178K}.

When varying individual parameters around the fiducial model, we (unsurprisingly) find that $f_{\rm III}$ (the fraction of pristine gas which forms stars in an episode of Pop III star formation) 
is the most 
important for setting the overall abundance of Pop III stars. Generally, we find that a significant (though subdominant)
amount of Pop III star formation continues to $z\lesssim6$. The behavior of the Pop III SFRD at low redshift 
depends strongly on the critical metallicity for metal-enriched star formation, $Z_{\rm crit}$ and the mass below which 
ionization feedback operates, $M_{\rm ion}$. For our fiducial model, it appears that the Pop III SFRD is not strongly 
decreasing with time at $z\approx 6$, however lowering $Z_{\rm crit}$ or raising $M_{\rm ion}$ begins to quench Pop III
star formation. While these parameters are the most important at lower redshifts, we point out that at high redshift 
the streaming velocity has a large impact. Increasing from $v_{\rm bc} = 1 \sigma$ to $v_{\rm bc} = 3\sigma$, substantially 
delays Pop III and metal-enriched star formation.  Another important result is that at $z \gtrsim 15$ the Pop III SFRD 
does not change for most of our parameter variations (with the notable exceptions of $v_{\rm bc}$ and $f_{\rm III}$). 
This is largely because the fraction of the box which is ionized and/or metal-enriched is quite low at early times (see Figure \ref{ion_avg}).
We also point out that while increasing $t_{\rm del}$ from our fiducial model did not have a strong affect on the Pop III SFRD, it 
did dramatically change the metal-enriched SFRD and sets a characteristic redshift where there is a transition from the
dominance of Pop III to metal-enriched star formation.

Due to the importance of external metal enrichment at $z\lesssim 10$, we also explore simultaneously varying $Z_{\rm crit}$ and $f_{\rm bub}$. For a critical metallicity corresponding to dust cooling, $Z_{\rm crit} = 10^{-6} Z_\odot$, we find that 
increasing the size of the metal bubbles can greatly reduce the Pop III SFRD at $z\approx 6$. Increasing $f_{\rm bub}$, but keeping $Z_{\rm crit} = 3\times10^{-4}~Z_\odot$ does not have a strong impact on the Pop III SFRD. This is because the metals get spread out over larger bubbles leading the metallicities to drop
below the critical value.

Several caveats must be kept in mind when interpreting our results.
A shortcoming of the model is that the simulation boxes used are 3 Mpc across, while
the horizon for LW photons is much larger ($\sim 100$ Mpc). Thus, while the LW background intensity 
is computed self-consistently,
it may not be accurate due to missing sources in rare over-dense regions which would be found in much larger
simulation boxes. Future work combining smaller-scale models like those presented here with $\sim {\rm Gpc}$ 
semi-numerical simulations such as those from \cite{2012Natur.487...70V, 2013MNRAS.432.2909F} will be required to make accurate predictions of the overall
abundance of Pop III stars \citep[see also][for a subgrid prescription of minihalos in cosmic reionization]{2012ApJ...756L..16A}. We also note that our star formation efficiencies ($f_{\rm II}$ and $f_{\rm III}$) are 
taken as constants throughout. In reality, these quantities may have both redshift and halo mass dependences which 
needs to be taken into account. Future work making more detailed direct comparisons with numerical simulations
will be required to determine the most accurate parameterization of the star formation efficiencies.
When varying parameters, we have generally changed one at a time with respect to the fiducial model. 
Simultaneously altering multiple parameters could lead to interesting new qualitative behaviors (as we saw for varying $f_{\rm bub}$ and $Z_{\rm crit}$ simultaneously), which we defer to future work. We also note that we have assumed very efficient mixing of metals in the IGM and in star-forming halos.
It has been suggested that inefficient mixing may lead to Pop III star formation in pristine pockets within halos with previous star formation \citep[e.g.,][]{2013ApJ...775..111P, 2019ApJ...871..206S, 2019ApJ...870L...3H}. We intend
to add this to our models in future work.

There are a number of upcoming observations that will have the ability to probe Pop III stars in the 
early Universe. These include 21cm observations in the pre-reionization era, observations of Pop III pair-instability 
supernovae explosions, and stellar archaeology of local extremely metal poor stars. Given the importance of 3-dimensional
effects in metal enrichment and reionization at $z \lesssim 15$, the grid-based approach described here will be particularly 
useful for predictions which depend on Pop III star formation at lower redshift (e.g.,~SN explosions or stellar archaeology). 
Future work running to even lower redshifts will determine when Pop III star formation is completely quenched. 
Finally, we plan to use the models presented here to  track the growth of black hole seeds formed from the remnants 
of Pop III stars. This will lead to interesting predictions for gravitational waves produced by merging Pop III seeds observable with \emph{LISA} 
and potentially for X-rays emitted during accretion onto $\approx 10^{4-5}~M_\odot$ black holes observable with 
an observatory such as the proposed mission \emph{Lynx}. 

\section*{Acknowledgements}
The Flatiron Institute (FI) is supported by the Simons Foundation. 
The numerical simulations were run on the FI cluster, Rusty.
GLB acknowledges support from NSF (grant AST-1615955, OAC-1835509), and NASA (grant NNX15AB20G), and computational support from NSF XSEDE. 
ZH acknowledges support from NASA (NNX15AB19G) and computational support from NSF XSEDE and NASA Pleiades.
\bibliography{paper.bib}

\end{document}